\documentclass[a4paper,11pt]{article}
\usepackage{graphicx,subfigure}
\usepackage[english]{babel}
\usepackage{amsmath,amsfonts,amsthm}
\usepackage{setspace}
\usepackage{cite}
\usepackage[utf8]{inputenc}
\usepackage{url}
\usepackage[nottoc,numbib]{tocbibind}
\usepackage{wrapfig}
\usepackage{floatrow}
\usepackage{wasysym}
\usepackage{capt-of}
\usepackage{epsfig}
\usepackage{transparent}
\usepackage{pdfpages}

\usepackage{titlesec}

\usepackage[small,bf,hang]{caption}

\graphicspath{{./Figures/}}
\newcommand{\pow}[1]{\times 10^{#1}}

\begin{document}

\begin{titlepage}
    \begin{center}
        \vspace*{1cm}
        
        \Large
        \text{Studies Supporting PMT Characterization for the IceCube Collaboration:}\\ 
        \vspace{0.2cm}
        \Huge
        \textbf{Precise Photodiode Calibration}
        
        \vspace{2cm}
        \LARGE
        Honours Award in Sciences
        
        \vspace{0.3cm}
        \Large
        \textbf{Wouter Van De Pontseele}

		\vspace{1.5cm}
        
        {\transparent{0.75}\includegraphics[width=0.35\textwidth]{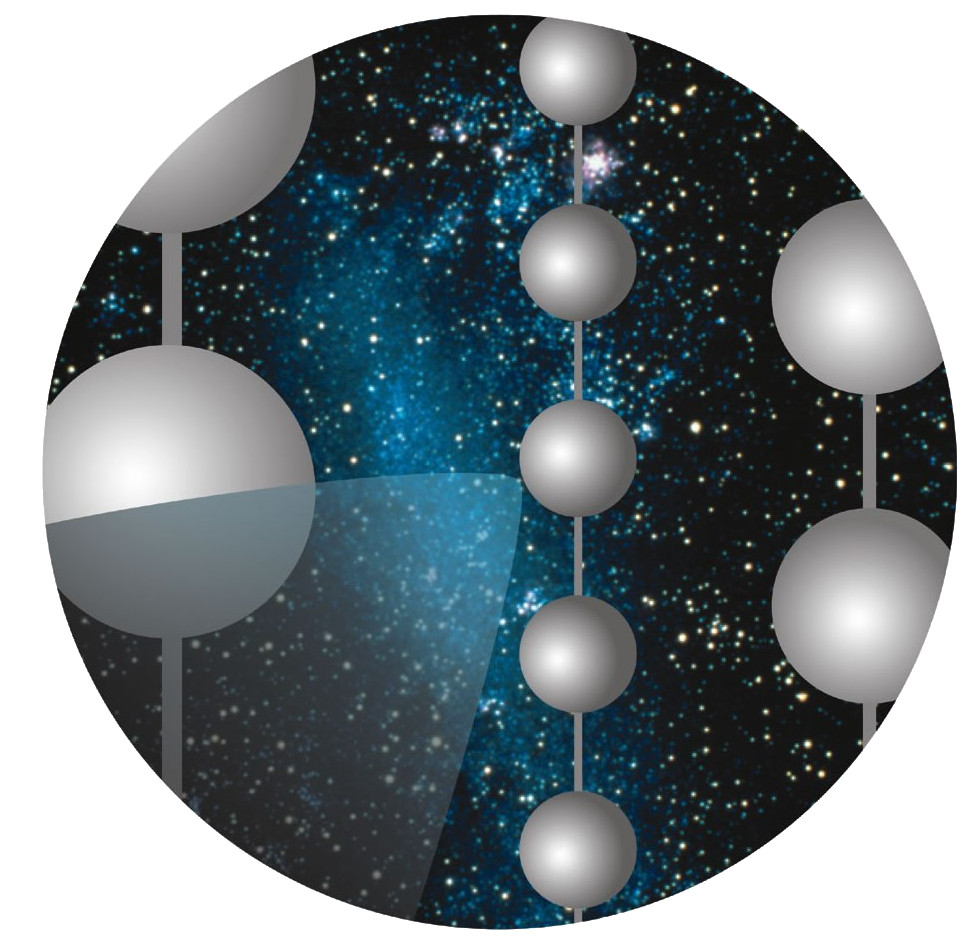}}
        
        \vspace{1.5cm}
        
        \begin{abstract} 
		A laboratory set-up has been developed to more precisely measure the DOM optical sensitivity as a function of angle and wavelength. DOMs are calibrated in water using a broad beam of light whose intensity is measured with a NIST calibrated photodiode. This study will refine the current knowledge of the IceCube response and lay a foundation for future precision upgrades to the detector. Good understanding of the photodiode readout is indispensable for the DOM calibration. Corrections to the photodiode measurements due to the amplifier circuit were investigated. To accomplish this, a general software structure has been added to the existing framework. Because the set of parameters in the source sector is still growing, modularity and a high level of automation were important objectives. The software features a large array of graphical tools to intercept problems at low level while the analysis can be easily adapted to the needs of foreseeable situations.
		\end{abstract}		
		
        \vfill
        
        \large
        Christopher Wendt - University of Wisconsin–Madison \\
        Prof. Dr. Dirk Ryckbosch - Ghent University\\
        \vspace{0.3cm}
        October 2015
        
    \end{center}
\end{titlepage}

\tableofcontents
\clearpage

\section{\label{sec:1}Introduction}
\subsection{A brief history of neutrino astronomy}

The neutrino was discovered by Cowan and Reines during an experiment based on inverse $\beta$-decay in 1956\cite{1956}. Soon after the observation, it was realised that the particle could be used as an ideal astronomical messenger. Neutrinos are leptons which have no electrical charge and almost no mass, as a result, they only interact due to the weak force. On their journey from the edges of the Universe they travel essentially without absorption or deflection by magnetic fields. This gives them an advantage over photons, that also interact electromagnetically. So, high-energy neutrinos may reach us unscathed  from cosmic distances, from the inner neighbourhood of black holes, and hopefully, tell us things about the furnaces where cosmic rays are born. Another thing to keep in mind is that cosmogenic neutrinos are signatures of hadronic interactions, while the origin of photons is more ambiguous. 

On the contrary, the above mentioned aspects of neutrinos make them very hard to detect. Huge detectors are needed to accumulate statistically significant data from experiments. Theoretically based estimates predict that the observation of potential cosmic accelerators such as gamma-ray bursts and quasars require a cubic kilometre-sized detector, at the time this was realised, around 1970, a daunting technical challenge. Given the detector's required size, it seemed logic to use large volumes of natural water and transform them to Cherenkov detectors that catch the light produced by neutrinos interacting with nuclei. 
The first early effort was the DUMAND Project (Deep Underwater Muon And Neutrino Detector Project). DUMAND was a proposed underwater neutrino telescope to be built in the Pacific Ocean, off the shore of the island of Hawaii five kilometres beneath the surface. It would have included thousands of optical sensors occupying a cubic kilometre of the ocean. Work began in about 1976, at Keahole Point, but the project was cancelled in 1995 due to technical difficulties. Despite this discontinuation, DUMAND paved the way for later efforts by pioneering many of the detector technologies in use today, and by inspiring the deployment of a smaller instrument in Lake Baikal, as well as efforts to commission neutrino telescopes in the Mediterranean Sea. The first telescope on the scale envisaged by the DUMAND collaboration was realized by transforming a large volume of the extremely transparent, natural deep Antarctic ice into a particle detector, the Antarctic Muon and Neutrino Detector Array (AMANDA). In operation from 2000 to 2009. It represented a proof of concept for the kilometre-scale neutrino observatory, IceCube\cite{halzen}.

\subsection{The neutrino spectrum}

\begin{figure}
\includegraphics[width=0.65\textwidth]{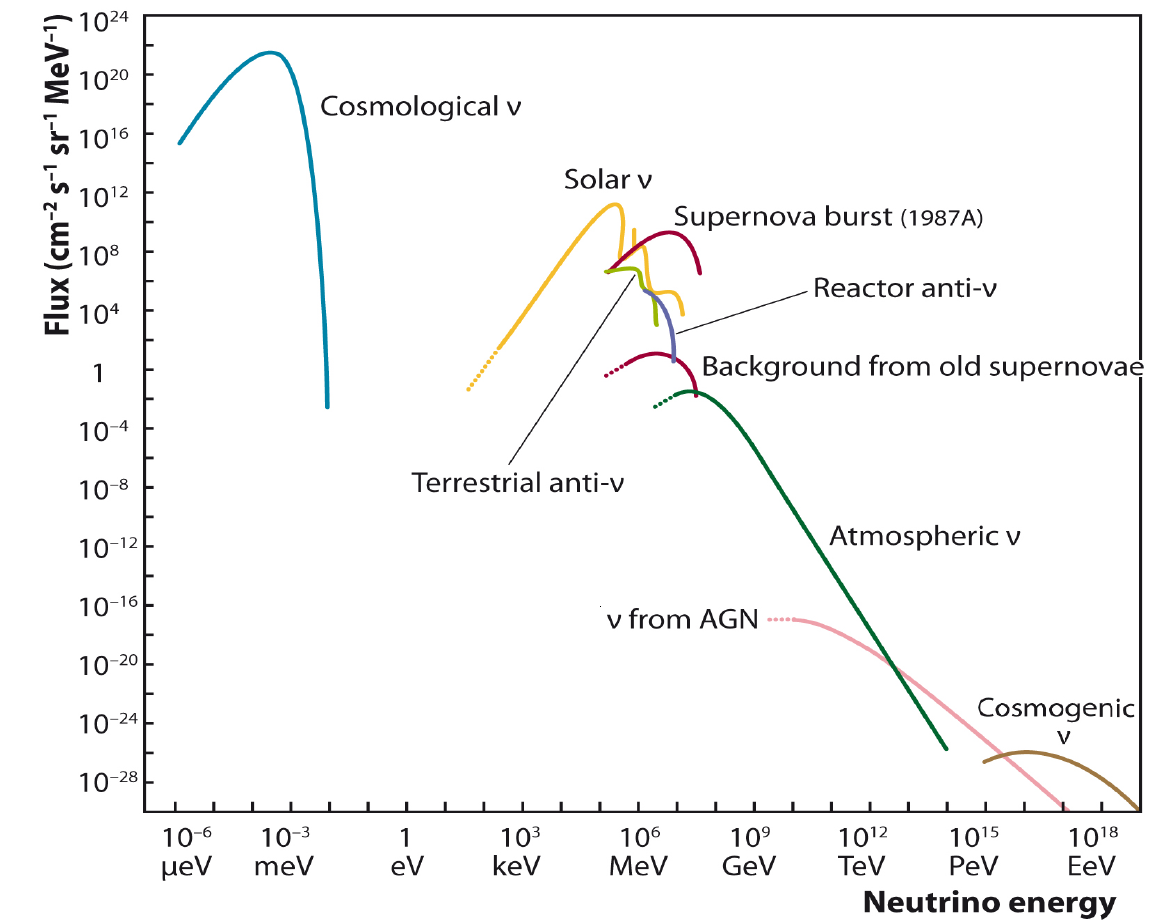}
\caption{Measured and expected fluxes of natural and reactor neutrinos. The energy range from keV to several GeV is the domain of underground detectors. The region from tens of GeV to about 100 PeV, with its much smaller fluxes, is addressed by Cherenkov light detectors; underwater and in ice. The predicted highest energies are only accessible with detectors one to three orders of magnitude larger than IceCube.~\cite{neutrinohistory}}
\label{fig:neutrinospec}
\end{figure}

The energy of a given neutrino, whether it is low or extremely high, gives us some clues about how and where it was produced. The different origins are depicted in figure~\ref{fig:neutrinospec}.

\paragraph{Low energy neutrinos} can be divided in two categories. The first is the cosmic neutrino background (C$\nu$B), a remnant of the Big Bang, which cannot be detected yet. Nevertheless, their characteristics can be predicted from cosmological models. The C$\nu$B has a very high flux but extremely low energy, they are indicated on figure~\ref{fig:neutrinospec} as cosmological $\nu$. 

The second category is dominant from KeV up to several MeV. Those neutrinos mainly find their origin in nuclear processes, such as the ones produced in nuclear reactors, the Sun or the centre of an exploding supernova. 

\subparagraph{High-energy neutrinos} come dominantly from the decay of pions and kaons, those heavier particles have their origin in collisions of comic rays with nitrogen and oxygen in the atmosphere. These neutrinos are therefore called ``atmospheric" in figure~\ref{fig:neutrinospec}. Their energy extends from a few MeVs up to tenths of a PeV. The first generations of Cherenkov detectors made clear that these atmospheric neutrinos would be a major background, at least for energies below 1 PeV, to searches for non-thermal astronomical sources where cosmic rays are accelerated. The spectrum of cosmic neutrinos from these sources extends to energies beyond those characteristic of atmospheric neutrinos.

\paragraph{Very-high-energy and Ultra-High-Energy (UHE) neutrinos} are categories that start around a few TeVs but can reach the PeV scale in rare occasions, the record of the highest neutrino ever captured is a muon neutrino of more than 2.6 PeV at IceCube in August 2015. Neutrinos which such a high energy cannot be created from within the solar system and are called cosmogenic. They can be divided in two kinds. The first kind of cosmogenic neutrinos were directly created from hadronic processes in or near the most extreme objects in our Universe, those powered by black holes and neutron stars.

The the second kind are decay products of pions produced by the interactions of cosmic rays with CMB photons. Cosmic rays above a threshold of around $4\pow{19}$ eV interact with the microwave background, introducing an cut-off feature in the cosmic-ray spectrum, the Greissen-Zatsepin-Kuzmin (GZK) cut-off~\cite{Greisen}. The value of this cut-off is estimated in appendix \ref{sec:A}. This phenomena limits the mean free path of extragalactic cosmic rays propagating in the microwave background to roughly 75 megaparsec. Therefore, neutrinos are the only probe of the UHE sources at longer distances. The GZK flux shares the high-energy neutrino sky with neutrinos from gamma-ray bursts and active galactic nuclei.

\subsection{Cherenkov detectors}

\begin{figure}
\includegraphics[width=0.4\textwidth]{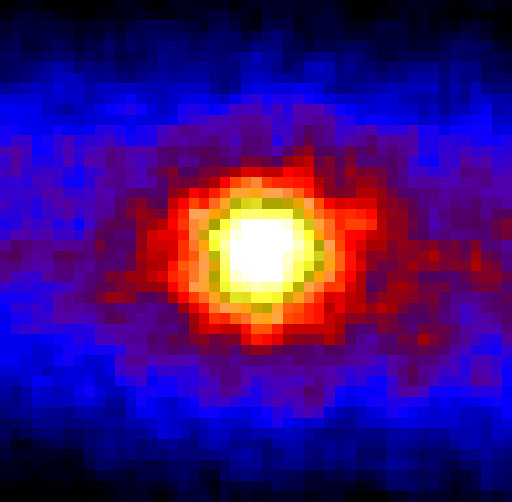}
\caption{Solar image using neutrinos captured with the Super-Kamiokande Cherenkov detector.}
\label{fig:neutrino-sun}
\end{figure}

However advanced the detector may be, most neutrinos will stream trough it without any kind of interaction. The few that interact with a nucleus create secondary muons or even hardronic or electromagnetic showers. The charged particles with enough kinetic energy radiate Cherenkov light.

Cherenkov light is radiated by charged particles moving faster than the speed of light in the medium; in ice, this is 75\% of the speed of light in a vacuum. A typical Cherenkov detector consists of some thousand photomultiplier tubes (PMTs). PMTs detect this blue and near-UV light. With a sufficient density of PMTs, neutrinos with energies of only a few MeV may be reconstructed. The water Cherenkov technique was first used in kiloton detectors, optimized for relatively low-energy (GeV) neutrinos. The two most successful first-generation detectors were the Irvine-Michigan-Brookhaven (IMB) and Kamiokande detectors. Both consisted of tanks containing thousands of tons of purified water, monitored with thousands of PMTs on the top and sides of the tank. Although optimized for GeV energies, these detectors were also sensitive to lower energy neutrinos; IMB and Kamiokande launched neutrino astronomy by detecting some 20 low-energy (MeV) neutrino events from supernova 1987A. Their success, as well as the accumulating evidence for the “solar neutrino puzzle”, stimulated the development of two second-generation detectors. Super-Kamiokande is a 50,000-ton version of Kamiokande, and the Sudbury Neutrino Observatory (SNO) was a 1,000-ton, heavy-water D$_2$O-based detector. Together, the two experiments clearly showed that neutrinos have mass by observing flavour oscillations in the solar and atmospheric-neutrino beams, thus providing the first evidence for physics beyond the Standard Model. Arthur B. McDonald, Takaaki Kajita were awarded the Nobel Prize in Physics (2015) for their contributions to SNO and Super-Kamiokande.

In summary, the field has already achieved spectacular success: neutrino detectors have “seen”
the Sun and detected a supernova in the Large Magellanic Cloud in 1987. Both observations
were of tremendous importance; the former showed that neutrinos have a tiny mass, opening
the first crack in the Standard Model of Particle Physics, and the latter confirmed the theory of
stellar evolution as well as the basic nuclear physics of the death of stars.

\subsection{The IceCube detector}

\begin{figure}
\includegraphics[width=.90\textwidth]{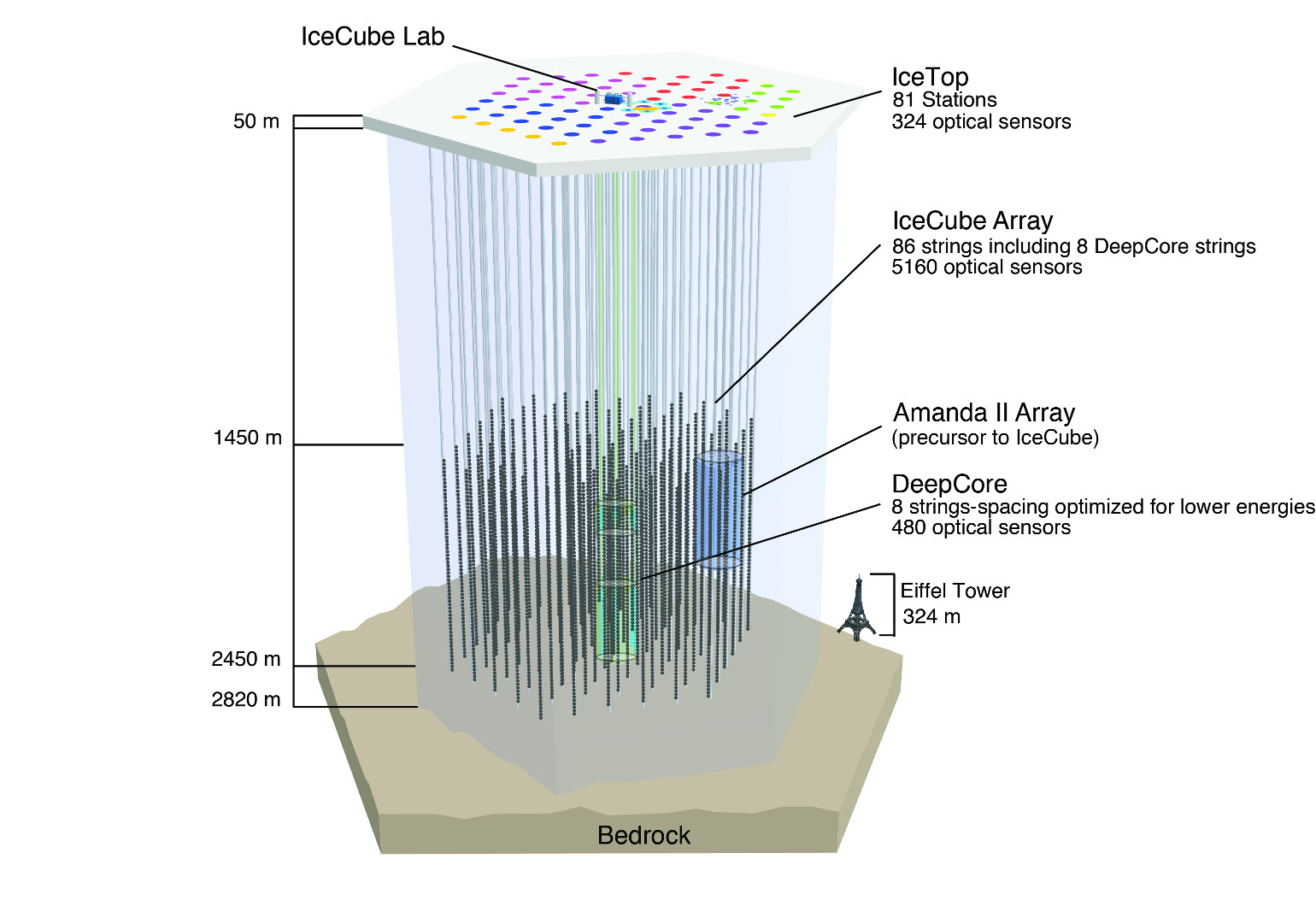}
\caption{Actual design of the IceCube neutrino detector with 5,160 optical sensors viewing a kilometre cubed of natural ice. The signals detected by each sensor are transmitted to the surface over the 86 cables to which the sensors are attached. IceCube encloses it’s smaller predecessor, AMANDA.}
\label{fig:detector}
\end{figure}

IceCube, the South Pole neutrino observatory, is a cubic-kilometre particle detector made of Antarctic ice and located near the Amundsen-Scott South Pole Station. It is buried beneath the surface, extending to a depth of about 2,500 meters. As hinted before, the construction of IceCube has been largely motivated by the opening of a new astronomical window. This allows us to learn about the most extreme places in our universe. 

There are two options for such a kilometre neutrino detector: liquid water or clear ice. As water can have a very good optical quality, it has the benefit of very good angular resolution. On the other side, decay of potassium and bioluminescence create bursts of background light that result in detector dead time. Another issue is the necessity to track the exact position of the PMTs because of the water currents.

Compared to water, natural deep ice has a shorter scattering length but a longer attenuation length which leads to an absorption length of more than 100m. Because the ice is extremely sterile, it also benefits from the absence of decays and bioluminescence~\cite{ice}. Appropriate reconstruction simulations showed that the PMTs can be placed a little further apart from each other in ice than in water.
Nevertheless, there is still a large scale water Cherenkov planned in the Mediterranean Sea, KM3NeT.

IceCube can measure neutrinos with energies above a few dozen GeV, which allows for measuring both the atmospheric and the extraterrestrial fluxes of neutrinos. Although atmospheric neutrinos are regarded as annoying background in most cases it is useful to calibrate the detector and compare results with other smaller Cherenkov detectors. The detector's major sensitivity is reached above the TeV scale, where the extraterrestrial flux is expected to become increasingly more dominant over the atmospheric neutrinos.

The in-ice component of IceCube consists of 5,160 digital optical modules (DOM), each with a photomultiplier tube and associated electronics. The DOMs are attached to vertical strings, frozen into 86 boreholes, and arrayed over a cubic kilometre from 1,450 meters to 2,450 meters depth. The strings are deployed on a hexagonal grid with 125 meters spacing and hold 60 DOMs each. The vertical separation of the DOMs is 17 meters. 
Eight of these strings at the centre of the array were deployed more compactly, with a horizontal separation of about 70 meters and a vertical DOM spacing of 7 meters. This denser configuration forms the DeepCore subdetector.  
IceTop consists of 81 stations located on top of the same number of IceCube strings. Each station has two tanks, each equipped with two downward facing DOMs. IceTop, built as a veto and calibration detector for IceCube, also detects air showers from primary cosmic rays. The surface array measures the cosmic-ray arrival directions in the Southern Hemisphere as well as the flux and composition of cosmic rays. All the parts described are illustrated in figure~\ref{fig:detector}.

There are several essential steps in the converting process of the messages from individual DOMs into light patterns that reveal the direction and energy of muons and neutrinos. The Cherenkov light is emitted by secondary particles after a neutrino interaction. Photomultipliers transform the Cherenkov light into electrical signals using the photoelectric effect. These signals are captured by computer chips that digitize the shape  of the current pulses. The information is sent to the computers collecting the data, first by cable to the IceCube Lab at the surface of the ice sheet and then via magnetic tape or, more recently, to the disk array storage. More interesting events are sent by satellite to the IceCube data centre in Madison, Wisconsin. 

Essentially, IceCube consists of 5160 freely running sensors sending time-stamped, digitized waveforms of the light they detect to the  surface. The local clocks in the sensors are kept calibrated with nanosecond precision. This information allows the scientists to reconstruct neutrino events and infer their arrival directions and energies. The complete IceCube detector observes several hundred neutrinos per day with energies above 100 GeV. The DeepCore array at the heart of IceCube identifies a smaller sample with energies as low as 10 GeV. It is the DeepCore part that is able to detect atmospheric neutrinos and creates the opportunity to study neutrino oscillations.

\section{\label{sec:2}Digital Optical Module (DOM)}

\begin{figure}
\includegraphics[width=0.8\textwidth]{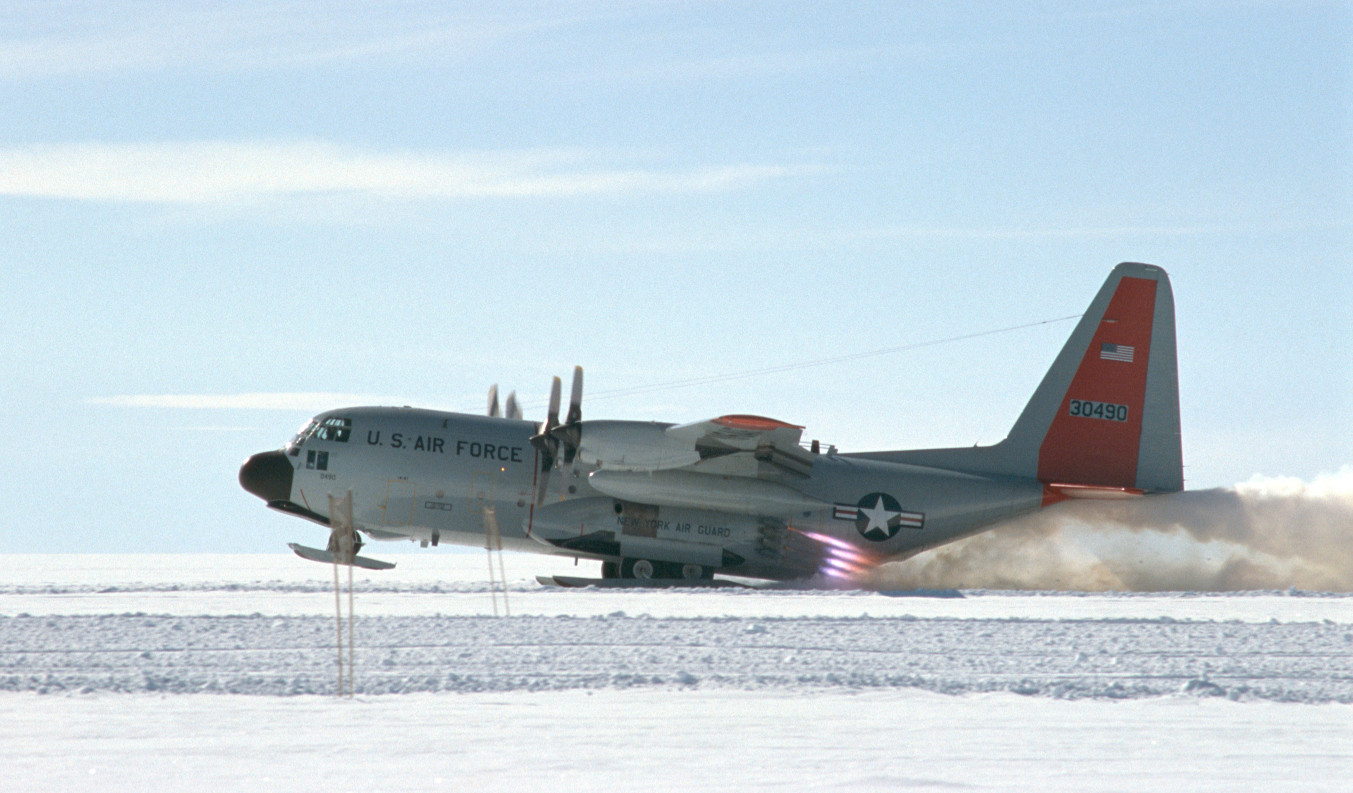}
\caption{An LC-130 rocket assisted take off.}
\label{fig:LC130}
\end{figure}

\begin{figure}
\includegraphics[width=0.8\textwidth]{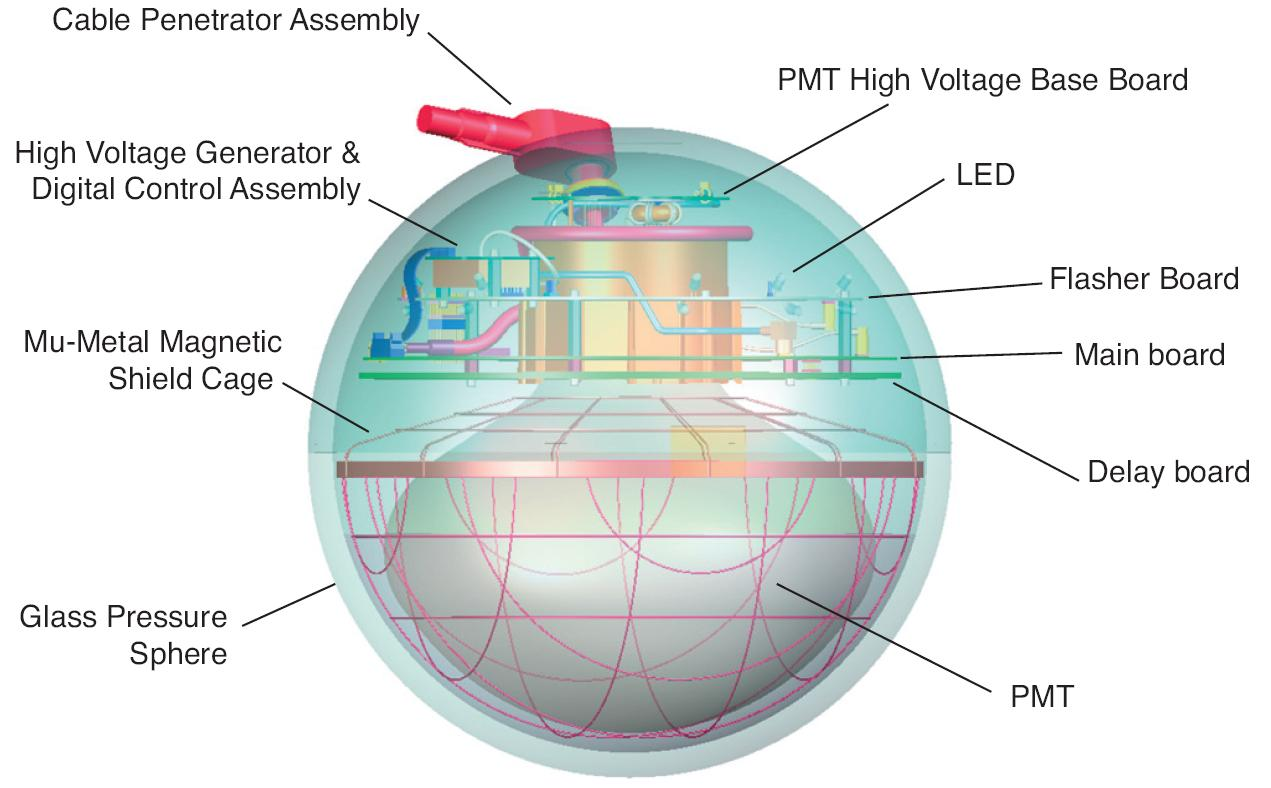}
\caption{Graph of cosmic microwave background spectrum measured by the FIRAS instrument on the COBE. This peaks at microwave wavelengths at a temperature of around 2.7K.Schematic drawing of a digital optical module \cite{eerstejaar}}
\label{fig:DOM}
\end{figure}

Each digital optical module is and integrated package containing a large photomultiplier, a high voltage circuit, LED flasher calibration board and a digital data acquisition system. The DOM mainboard is the core of the IceCube data acquisition system. It contains logic responsible for reading out, digitizing and buffering the PMT signals. The layout of the module is showed in figure~\ref{fig:DOM}. All internals are housed in a glass casing. 

The DOM development had to overcome challenging design problems. There were the extremes of high reliability in a remotely-deployed high pressure and low temperature environment. The borosilicate glass sphere encasing the DOM internals protects them from the immense pressures, up to 400 ATM on ice refreeze, and is selected for high UV transmission. To be more specific, the case is able to withstand pressures up to 70 MPa and transmits light with wavelengths longer than about 350nm. Besides that there was a stringent limit on the allowed radioactivity of the vessel to maintain a maximal dark noise rate of 500Hz. The electronics in it had to be operational from room temperature for testing down to $-55^\circ$C. The reliability requirement had to be on the same level as for satellites because the DOMs are totally inaccessible after deployment. To assure this, the boards and completed DOMs were subject to stringent testing. Prototype boards were subjected to HALT (Highly Accelerated Lifetime Test) cycling, including high and low temperatures, rapid temperature cycling, and high vibration levels. Thermal imaging was also used to check for hot spots. All of the production boards were subjected to HASS testing, a less-stressful version of HALT. Ninety-eight percent of the DOMs survive deployment and freeze-in completely; another 1\% are impaired, but usable (usually, they have lost their local coincidence connections). Post-freeze-in reliability has been excellent, the estimated 15-year survival probability is 94\%.

Since all the fuel has to be flown in on LC-130's, the polar version of the C-130 cargo plane with ski-equipped landing gear, the DOMs also had to be as low power as possible. For entertaining purposes, the plane can be found in figure~\ref{fig:LC130}. With the currently used configuration, each digital optical module consumes about 3.5 W.

Another requirement was a huge dynamic range: from single photon hits, caused by passing cosmic-ray muons, up to tens of thousands of photons from immense electromagnetic showers. The time tagging of events had to be of nanosecond precision. The system response to a single photo-electron (SPE) is a pulse with an average amplitude of about 10 mV and a width of 5 ns. The timing is slightly sensitive to where the photoelectrons hit the photo-cathode; photons striking the edges of the PMT are recorded, on average, 3 ns later than those reaching the centre. Their time resolution is also worse.

Last but not least, The DOMs had to be cost effective and reliably producible in quantities of a few thousand units \cite{DOM}.

Figure~\ref{fig:DOM} shows the main parts of the optical module, which has a total diameter of 35 cm. The largest part is occupied by a 25 cm photomultiplier tube from Hamamatsu and associated electronics \cite{PMTpaper}. The amplifying section has ten dynodes and runs at a gain of $10^7$ at $1500$ V. A mu-metal magnetic shield reduces the magnetic field of the earth with a factor two. The PMT is optically coupled to the pressure vessel using an optical gel and is sensitive to 350-650 nm photons. The high voltage of $1300-1500$ V is created with a Cockroft-Walton power supply and can be adjusted from the south pole station. Each DOM also contains 13 LED's used for photonic calibrations.

Along each string of modules, a cable connects all of them with the IceCube surface counting house. The connection between the cable and the DOM is possible trough the cable penetrator assembly. The cable incorporates local coincidence circuitry, whereby DOMs communicate with their nearest neighbours. IceCube DOMs have several operating modes for the local coincidence circuitry. Until early 2009, IceCube ran in “Hard Local Coincidence” mode. In this mode, the DOMs only saved data when two nearest neighbour or next-to-nearest-neighbour DOMs saw a signal within a $\mu$s coincidence window. The hit rate in this mode depends on a DOM’s depth, through both the muon flux and the optical properties of the ice, but is typically 3 to 15 Hz.

In early 2009, IceCube started taking data in “Soft Local Coincidence” mode. In addition to the complete data for coincident hits, a more selective part of the data was sent to the surface for isolated hits. These hits are recorded at the PMT's dark rate, typically 350 Hz. Although most of these hits are noise, they are useful in many analyses.

\section{\label{sec:3}In situ hardware calibration techniques}

Determining the time and amplitude of an observed light pulse requires careful calibration of the instrumentation. IceCube uses a variety of methods to ensure this. The primary timing calibration is “RapCal”: Reciprocal Active Pulsing. RapCal timing calibrations are performed automatically every few seconds. During each calibration, the surface electronics send a timing signal down to each DOM, which waits a few $\mu$s until cable reflections die out, and then sends an identical signal to the surface. The surface and DOM electronics use identical DACs and ADCs (Analog-to-digital converters) to send and receive signals, so the transmission times in each direction are identical. Even though the $3.5$km cable transmission widens the signals to $\approx 1 \mu$s, the transmission time
is determined to less than 3 ns. Other timing calibrations measure the signal propagation delay through the PMT and electronics. Each main board includes a UV LED, which may be pulsed on command. The LED pulse current is recorded, along with the PMT signals. The difference determines the PMT transit time, plus the delay in the delay line and other electronics. Amplitude calibrations are also done with the On-Board LED. It is flashed repeatedly at low intensity. A charge histogram is accumulated and sent to the surface, where it is fit to find the single photo-electron peak. This is done for a range of high voltages, and the high voltage is set to give $10^7$ PMT gain. These calibrations are extremely stable over time periods of months. Each DOM also contains a ‘flasher’ board with 12 LEDs mounted around its edges. These LEDs are used for a variety of calibrations, measuring light transmission and timing between different DOMs. The multiplicity of LEDs is particularly useful for linearity calibrations. The LEDs are flashed individually, and then together, providing a ladder of light amplitudes that can be used to determine the saturation curve. 

\begin{figure}
\includegraphics[width=0.8 \textwidth]{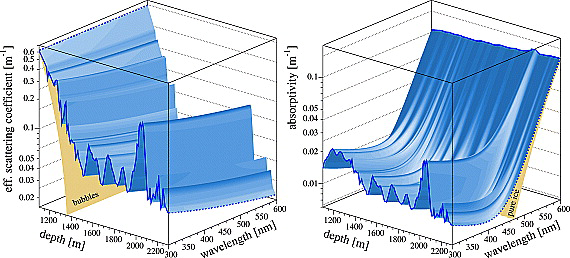}
\caption{Maps of optical scattering (left) and absorption (right) for deep South Pole ice. The depth dependence between 1100 and 2300 m and the wavelength dependence between 300 and 600 nm for the effective scattering coefficient and for absorptivity are shown as shaded surfaces, with the bubble contribution to scattering and the pure ice contribution to absorption superimposed as (partially obscured) steeply sloping surfaces. The dashed lines at 2300 m show the wavelength dependences: a power law due to dust for scattering and a sum of two components (a power law due to dust and an exponential due to ice) for absorption. The dashed line for scattering at 1100 m shows how scattering on bubbles is independent of wavelength. The slope in the solid line for absorptivity at 600 nm is caused by the temperature dependence of intrinsic ice absorption \cite{ice}.}
\label{fig:icespecs}
\end{figure}

Further more, mounted between two DOMs on a cable, there are extra modules containing a 337-nm $N_2$ laser. The laser beam is shaped to emit light in the shape of a Cherenkov cone, forming a reasonable approximation to a cascade. The light output is well-calibrated, and an absorber wheel allows for variable intensities. Although the 337-nm light does not propagate as far as typical Cherenkov radiation (peaked around 400 nm), it provides a reasonable simulation of cascades up to PeV energies. Why 400 nm photons propagate further can be deduced from a combination of the scattering and absorbtion characteristics of the ice as can be seen in figure~\ref{fig:icespecs}.

\section{\label{sec:4}Experimental set-up for ex situ calibration}

\begin{figure}
\includegraphics[width= \textwidth]{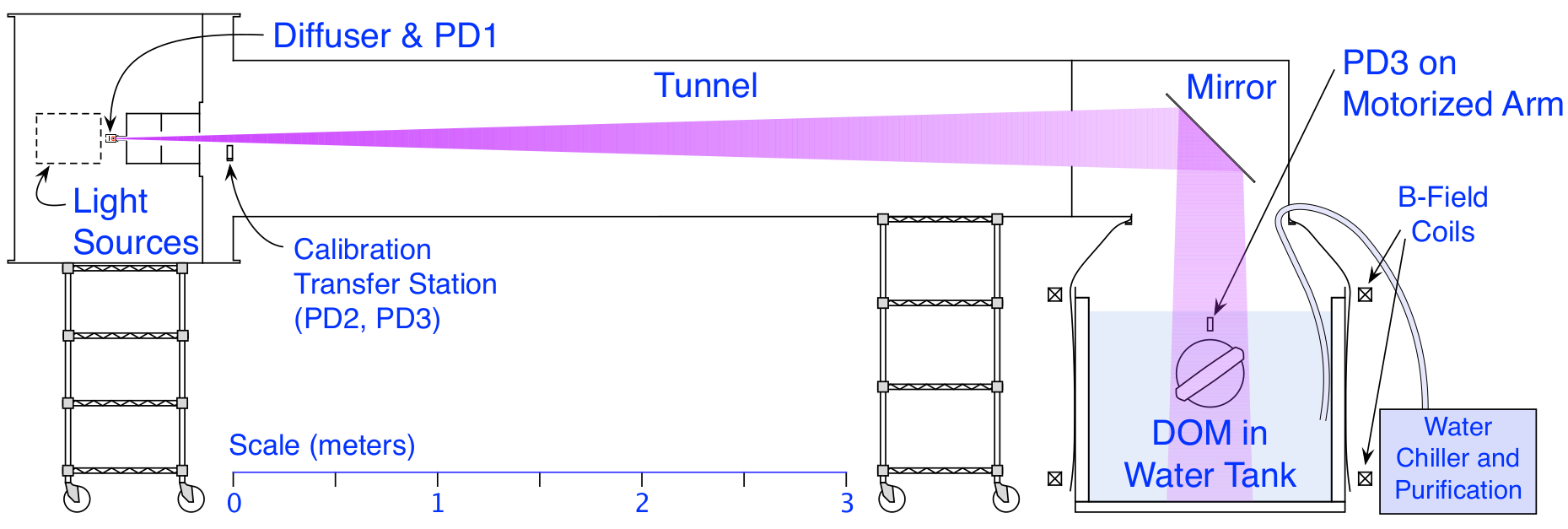}
\caption{Illumination system and water tank. Photodiodes are placed as shown and used as described
in the text: (PD1) Relative measurement of beam intensity for bright or dim source settings; (PD2) NIST calibrated, used to calibrate PD3 when temporarily mounted at transfer station; (PD3) Direct measurement of beam flux at DOM for bright source settings.}
\label{fig:setup}
\end{figure}

\begin{figure}
\includegraphics[width= \textwidth]{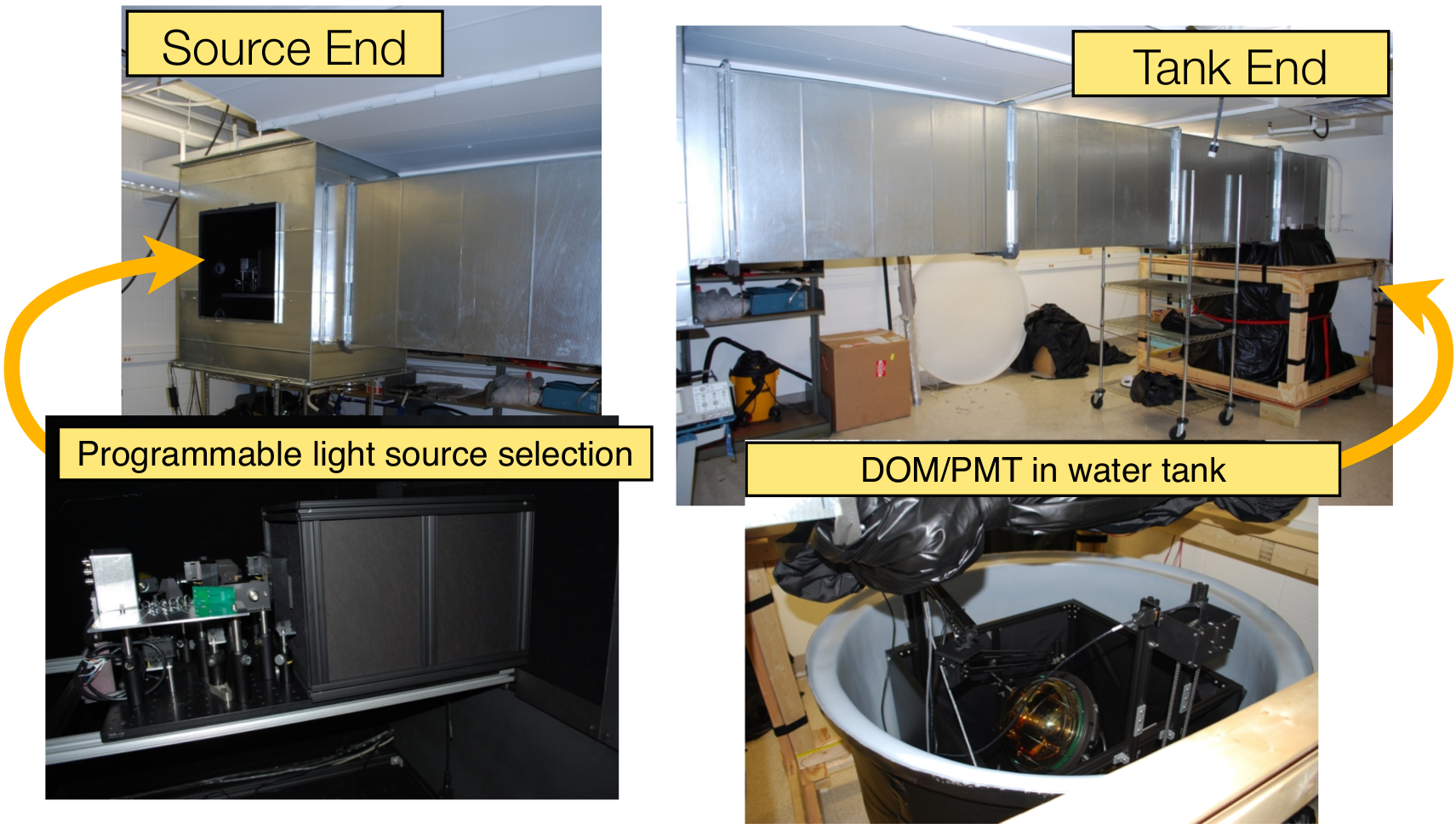}
\caption{Photographs of the acual set-up as sketched in figure~\ref{fig:setup}. A more detailed view of the illumination system can be found in figure\ref{fig:sources}.}
\label{fig:setuppic}
\end{figure}

\begin{figure}
\includegraphics[width= 0.4\textwidth]{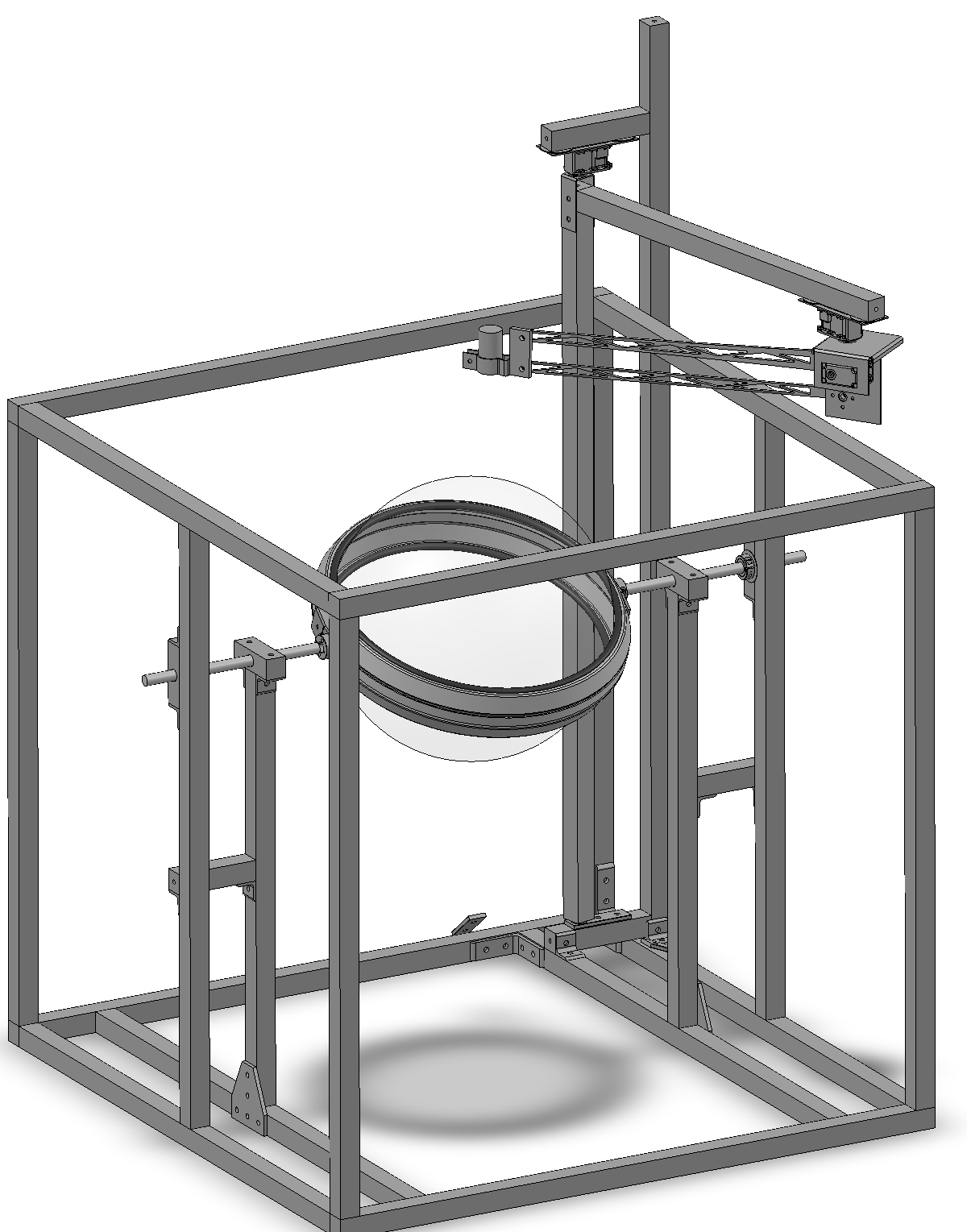}
\caption{Schematic of cage where photodiode 3 (PD3) and the DOM are submerged. Because the the orientation of incident light cannot be adjusted, there is equipment installed to position PD3 and change the angle of the DOM with respect to the light.}
\label{fig:rotation}
\end{figure}

\begin{figure}
\includegraphics[width= \textwidth]{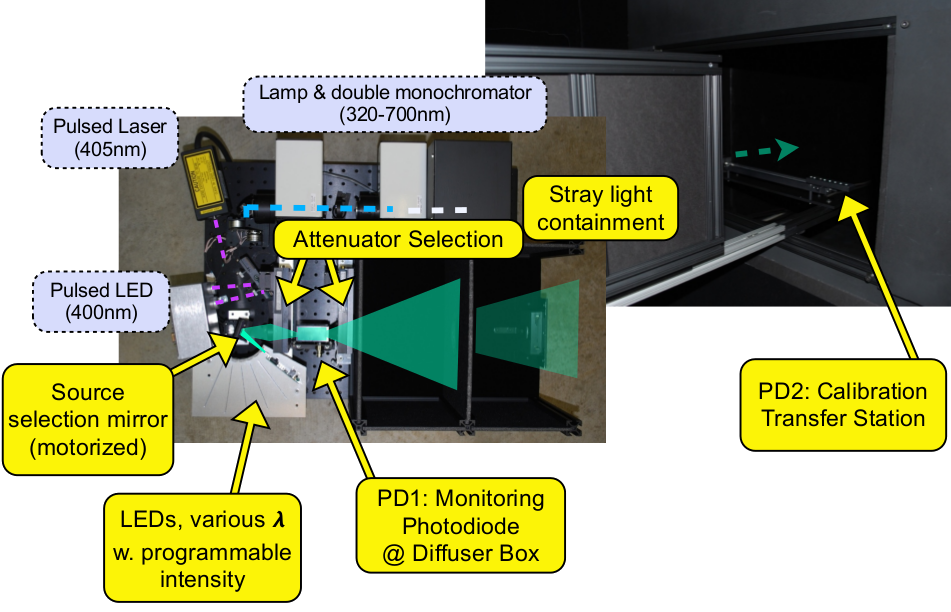}
\caption{The currently installed light sources include: a pulsed diode laser (405 nm), pulsed LEDs (400 nm),
continuous LEDs (370 nm, 400 nm, 450 nm), and lamp with monochromator (320-700 nm). Beam
intensity entering the diffuser box can be varied over a wide range by means of LED current,
repetition rate of pulsed sources, neutral density optical filters, and a simple shutter.}
\label{fig:sources}
\end{figure}

\begin{figure}
\includegraphics[width= \textwidth]{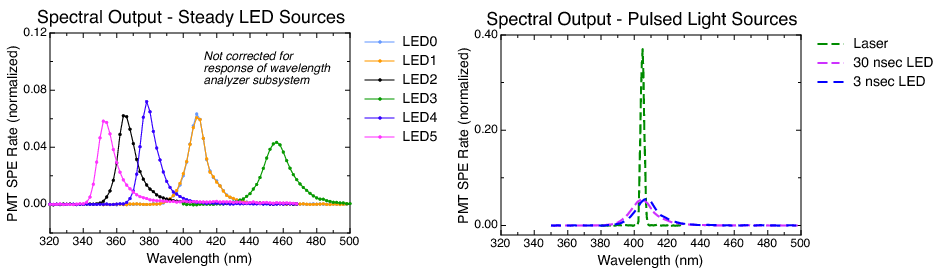}
\caption{Spectral output characteristics of the LEDs and the laser.}
\label{fig:sourcesgraph}
\end{figure}

Depending on the energy of the neutrino and the distance from secondary particle tracks, PMTs can be hit by up to several thousand photons within a few hundred nanoseconds. The number of photons per PMT and their time distribution is used to reject background events and to determine the energy and direction of each neutrino. The detector energy scale was established from previous lab measurements of DOM optical sensitivity, then refined based on observed light yield from stopping muons and calibration of ice properties. A laboratory set-up has been developed to more precisely measure the DOM optical sensitivity as a function of angle and wavelength. DOMs are calibrated in water using a broad beam of light whose intensity is measured with a NIST (National Institute of Standards and Technology) calibrated photodiode. This study will refine the current knowledge of the IceCube response and lay a foundation for future precision upgrades to the detector.

The lab set-up is designed to measure the single photon detection efficiency of a DOM when illuminated in a similar way as from neutrino events in IceCube, where light typically travels 5-200 m from its source before possible detection. Because the DOM diameter is only 35 cm, at such distances it can be accurately described by its efficiency to detect photons arriving from a particular angle, averaging over all possible points of arrival at the photocathode. This situation is mimicked in the set-up by using a uniform light beam from a source 6.4 m away, as can be seen in figure~\ref{fig:setup}.

In IceCube, each DOM has its PMT facing downward, and the sensitivity does not vary significantly with rotation around this vertical axis. The sensitivity does depend on the polar angle of illumination relative to the DOM axis, so the lab set-up includes a motorized mounting shaft for inclining the DOM axis relative to a fixed vertical beam. This is illustrated in figure~\ref{fig:rotation}. The DOM is immersed in water in order to closely model reflection and refraction effects occurring at the optical interface as in the case that the DOM would be embedded in ice. Effects of PMT gain, discriminator threshold and electronics calibration are accounted for by using the same software and procedures as in standard IceCube operations. 

The set-up as sketched in figure~\ref{fig:setup} is made up of two major parts: the illumination system and the water tank. The corresponding parts are photographed in figure~\ref{fig:setuppic}. There are several sources of light available. The spectral output of the LEDs and the laser is depicted in figure~\ref{fig:sourcesgraph}. Any of those can be directed to shine towards a diffuser box as indicated in figure~\ref{fig:sources}. The diffuser acts as an integrating sphere and produces a homogeneous beam that does not depend on the kind of source that was used. The box has two outputs. The main output goes to the water tank after being deflected by a mirror. Photodiode 1 (PD1) is mounted on a secondary port of the source diffuser and is used to precisely measure changes in beam intensity. The beam geometry is defined by a series of apertures in the tunnel, creating a uniform circular disc of light with a 40 cm diameter at the DOM position. 

When sources are operated in a bright mode, PD3 is used to precisely measure the light reaching the DOM, as well as to test uniformity and boundaries of the beam cross section. When dimmer sources are used, PD3 is unable to measure the light flux any more. It is is moved out of the way and PD1 measures the intensity now. The PD1 measurements can be converted to the intensity at the DOM~\cite{Deliapaper}. Closer to the diffuser output, a calibration transfer station allows temporary mounting of a photodiode side by side with a NIST calibrated one. Photodiode currents are measured with custom pre-amplifier and ADC boards. 

The tank water is continuously pumped through an external cooler and water purification system. Maintaining the temperature around $5^{\circ}$C avoids introducing humidity to the tunnel
and source region, where it might affect electronic or optical elements. The low temperature also
reduces the DOM dark noise rate and discourages bacterial growth in the water.

Surrounding the water tank, Helmholtz coils are arranged for separate control of the magnetic
field in each direction. These coils cancel the ambient field in the lab and create a field relative to
the DOM axis. The field can be generated to mimic that experienced at the South Pole for all possible rotation angles of the DOM in the set-up. A blackout curtain hangs from the open end of the tunnel and is secured around the water tank by a belt.

All system elements are controlled and monitored by Python scripts running on a standard
IceCube data acquisition computer (DOMHub)~\cite{domhub}.

\section{\label{sec:5}Measurements}

\subsection{The importance of PD1}

With PD3 installed in the water tank and one of the light sources turned on, a PD3 current around 100 fA can be obtained if the source settings are bright enough. Such a current enables measurements with a precision of 0.5\%. This gives a direct measurement of the beam flux in the tank, typically $10^6$ photons/cm${}^2$ /sec at high brightness settings. However, the PD3 current signal cannot be precisely measured at the much lower brightness settings needed for DOM sensitivity calibration. For beam fluxes below 100 photons/cm${}^2$ /sec, PD1 is used. This photdiode is located much closer to the source. The larger PD1 signal is still proportional to the beam flux, but with an scale factor that has been calibrated against the direct PD3 measurement~\cite{Deliapaper}. For this purpose a bright PD3 measurement and a PD1 measurement has been done simultaneously, yielding the beam flux scale factor:
\begin{equation*}
\text{Flux} = 0.0226 \: \text{photons/cm}^2 \text{/sec} \cdot \text{(PD1 current / fA)}
\end{equation*}
Since PD1 and PD3 were both observing outputs of the source diffuser box, this scale factor applies equally when the source is operated in dim mode. The switchable gain preamp used with PD1 facilitates measurement of both bright and dim beams with good precision (Table~\ref{table1}). For example, a dim beam with 10 photons/cm${}^2$ /sec at 400 nm gives DOM count rates up to 1000 Hz, with PD1 currents around 500 fA. Such dim measurements are then made for each polar angle and wavelength and used to calculate the final DOM photon sensitivity.

\begin{table}
\centering
\caption{Approximate response of PD1, PD3 and DOM for bright and dim beams.}
\label{table1}
\begin{tabular}{c|c|c|c|c}
Source setting & Photon flux        & PD1    & PD3     & DOM         \\ \hline
Bright    \rule{0pt}{3ex}      & $10^6$ /cm$^2$/sec & 50 nA  & 100fA   & saturation  \\
Dim            & $10$ /cm$^2$/sec   & 500 fA & too low & 100-1000 Hz \\ 
\end{tabular}
\end{table}

\subsection{The switchable gain preamp used with PD1}

\begin{figure}
\includegraphics[width= \textwidth]{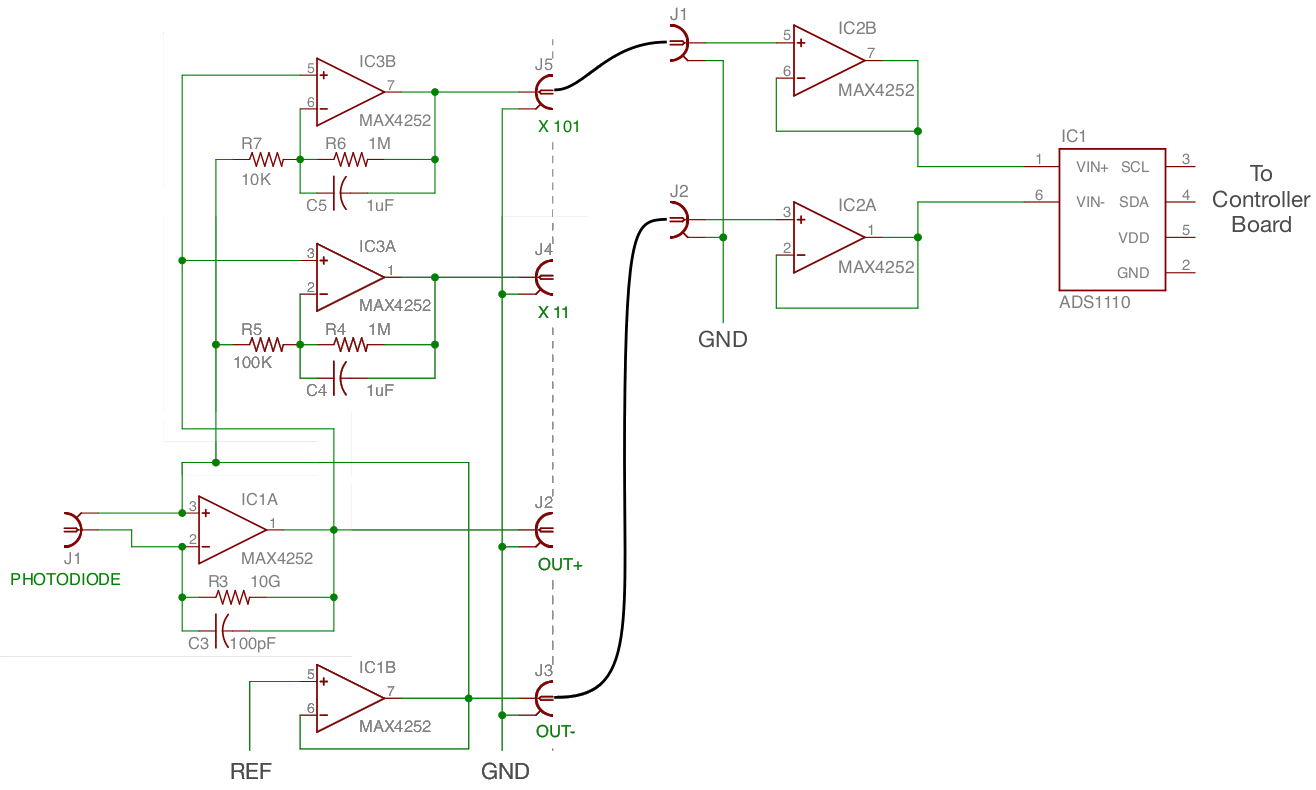}
\caption{The preamp circuit that connects the output of PD1 to the controller board.}
\label{fig:pd1circ}
\end{figure}

From the previous discussion it became clear that PD1 needs to cover a very large range of intensities while maintaining small errors. To accomplish this goal an advanced pre-amplifier circuit was added to the ADC board. The schematic can be found in figure~\ref{fig:pd1circ}. The amplifying process consists of three stages, all of those use the principle of operational amplifiers. A short introduction to operational amplifiers is found in appendix~\ref{sec:B}. The important point in the first amplification is that in the gain is proportional to the resistor that is used in the circuit. In the first stage in the amplifying process a gain resistor could be chosen from a set of six resistors ranging from 10 k$\Omega$ up to 1 G$\Omega$. The second gain parameter was the channel. Two channel options were implemented in the software: \textit{ch0} represented the normal signal, while \textit{ch1} amplified the original signal by a factor 101. The third parameter in the fine tuning of the total gain was called the ADC-gain and generated by two separated low-noise, low-distortion commercial operational amplifiers (MAX4252). Each of those had 1,2,4 and 8 as possible gains, all of them were implemented in the code but during the effective measurements only options 1 and 8 were used to limit the parameter space. 

The goal of the project was to determine relative errors on the different gain combinations. Those ratios will then be used as correction factors in the further experiment, as described in section~\ref{sec:4}. 

\subsection{The parameters of the source sector}

The incoming light can be regulated by adjusting the following parameters:

\paragraph{Source type} The installed sources were described in figure~\ref{fig:sources}. For this project only LEDs were used because a steady source was required and there was no need for a specific wavelength.  

\paragraph{LED number} At the moment of the experiment, there were five LEDs installed with slightly different characteristics (see figure~\ref{fig:sourcesgraph}). Number 0 was used in almost all cases, occasionally LED 3 was used because of its slightly higher brightness level. This made it possible to accumulate more data for the smallest gain settings. Number 3 was also used to test the source independency of certain behaviour of the measurements.

\paragraph{Brightness range} The circuit that drives the LEDs had two parameters of which the first was the range: bright or dim. The bright range was about a factor 40 brighter than de dim range.

\begin{figure}
\includegraphics[width= 0.65\textwidth]{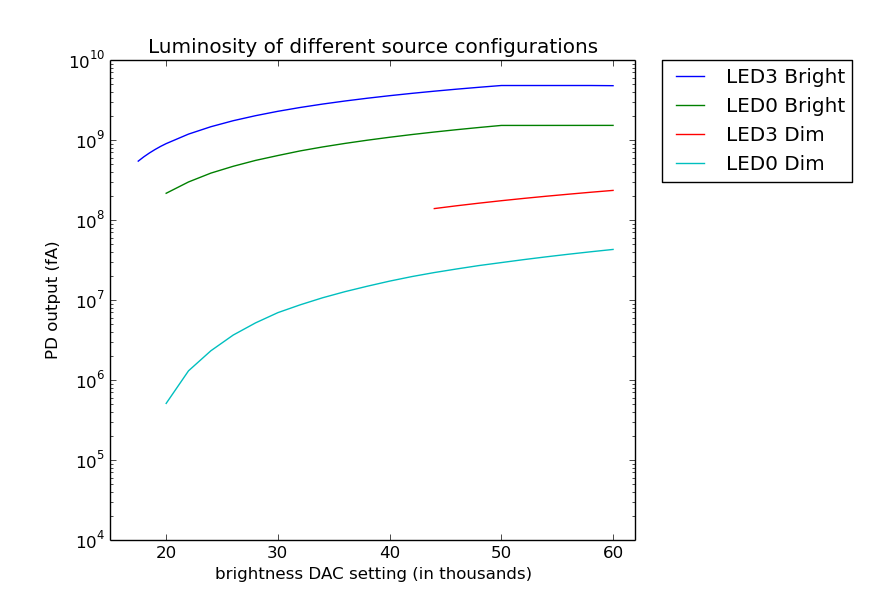}
\caption{Behaviour of the LED output measured with PD1 for the different settings used in the further analysis. All measurements showed in this plot were made without an extra filter in between the LED and PD (filterwheel 1 on position 1). For low DAC settings, the behaviour is non-linear.}
\label{fig:sourcesself}
\end{figure}

\paragraph{Brightness Digital-to-Analog Converter (DAC) setting} This second source parameter accepts values from 0 up to 65535. The effective luminosity produced by the LEDs was not linear and in practice there was a threshold around 17k and an upper bound around 60k. In the bright range the LED output became constant when values over 50k were used. The combination LED 0 and 3, dim/bright range and DAC setting already gave a pretty good coverage of the possible luminosities as is made visual in figure~\ref{fig:sourcesself} 

\begin{table}
\centering
\caption{Attenuation measurements of different filterwheel1 settings.}
\label{Tcoef}
\begin{tabular}{c|c}

filterwheel1 position & transmission coefficient \\ \hline
1                     & 1                        \\
4                     & 0.002599                 \\
5                     & 0.000392                 \\
6                     & 0.000247                 \\ 
\end{tabular}
\end{table}

\paragraph{Filterwheels} The last piece of hardware that makes it possible to further tune the amount of light that reaches PD1 is the so called filterwheel1 (fw1). Positions 1,4,5 were used for measurements. There is also a filterwheel2 located behind PD1, this was used on position 6 to prevent bright light from reaching the DOM PMT. position 2 and 3 of the filterwheels are reserved for bandpass filters and are not used. The T coefficients of positions 1,4,5 and 6 can be fount in table~\ref{Tcoef}. This extra parameter enables to go to the very lowest brightness levels. The calibration of the the highest gains to amplify the PD1 signal on these brightness levels is extremely important since it matches the DOM south-pole conditions more accurately than brighter settings.

If we group the gain settings and the source settings together, one measurement can be defined by:
\begin{itemize}
\item Source settings:
\begin{itemize}
\item source type (LED)
\item LED number
\item Brightness range
\item Brightness DAC setting
\item fw1 and fw2 position
\end{itemize}
\item Readout settings:
\begin{itemize}
\item Resistor gain (10k, 100k, 1M, 10M, 100 or 1G)
\item ADC gain for \textit{ch0}
\item ADC gain for \textit{ch1}
\end{itemize}
\end{itemize}

To keep things comprehensible, the the ADC-gain for both channels will always be the same during the data taking. More specific both \textit{ch0} and \textit{ch1} will have ADC-gain 1 or 8.

A main concern during the automation process is that the gain setting under investigation has to be matched with the right range of source settings. This restriction arises because of the limiting range of the counting rate of the ADC controller board. The digital readout was an integer value between 0 and 32767.

\subsection{Structure of a measurement}

\begin{figure}

\includegraphics[width=\textwidth]{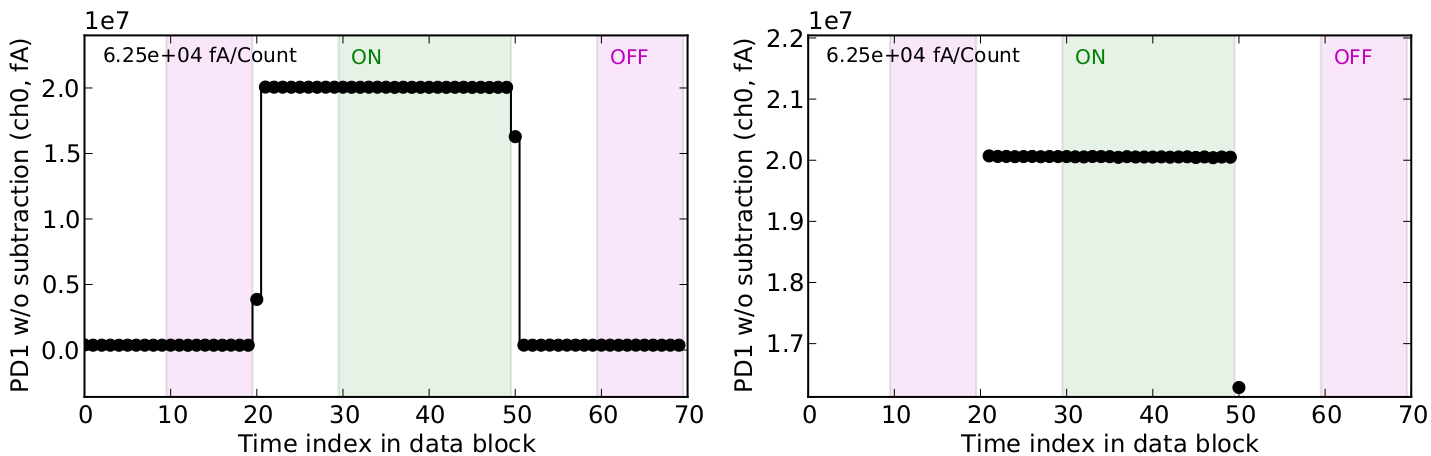}
\caption{Intermediate result of an example of a PD1 measurement. This is one block with specific source and gain settings. White coloured blocks indicate \textit{settling\_time}, red ones are \textit{off\_time} to determine the background off-set and the green part is \textit{on\_time}. The right part is just zoomed in. }
\label{fig:time}
\end{figure}

Once the output parameters are known there are still a few parameters that need to be set considering the structure of a measurement. Figure~\ref{fig:time} will serve as an example to clarify their meaning.

\begin{itemize}
\item \textit{settling\_time}: buffer period before on/off data-taking (10 s by default).
\item \textit{on\_time}: time that the source is on and that the data is kept for analysis (20 s by default).
\item \textit{off\_time}: time that the source is off and that the data is kept for analysis (10 s by default). 
\end{itemize}

Those parameters were only changed in a try to resolve some peculiar behaviour that will be discussed in section~\ref{sec:8}. Most of the time, there default value was used. With those parameters, one measurement took exactly 70 seconds.

\section{\label{sec:6}Automatization of the measurement process}

\subsection{General remarks on the program}

From the preceding, it becomes clear that there are al lot of source parameters to play with. Even more challenging is the number of different gain combinations that has to be investigated. To calculate the relative errors on the different gain combinations, one has to scan over all the combinations of gain settings and try to find as much as possible cases were different gain settings can be used to measure the brightness of a source with identical source parameters. The ratio of the lowest and the highest gain is: 
\begin{equation}
\frac{Gain_{highest}}{Gain_{lowest}} = \frac{1\text{G} \cdot 101 \cdot 8 }{10\text{k} \cdot 1 \cdot 1} \approx 8.08 \pow{7}.
\end{equation}
At the moment we neglect possible error on the gain settings because they are believed to be small, according to manufacturers supplied data. As this number is far larger than the counting range of the ADC board, it will be impossible to measure even half of the combinations with a constant source setting. To accumulate data and investigate source dependencies, the goal is to measure all gain settings for the whole range of possible brightness levels that results in an output count rate that lies in the interval of the ADC board.
From this, we can already anticipate on the fact that we will need to combine intermediate results if we want to cover the whole range of gain ratios.

Another remark is that each single measurement takes 70 seconds, and has to be done at least 3 times (cycles) to be sure that the result is time independent and that the source is stable. As a result, the total scan over all the options takes a few days up to a full week. 

Those two arguments ask for an automated programme that runs on a dedicated hub that is installed next to the experimental set-up and is configured exactly the same as the ones used on the south-pole. Another requirement was that the analysis of the data could be done on the same locally installed computer. For this last task we quickly ran into problems because of the limiting analysis capacities of the standard IceCube configured software set-up. A  compromising solution was found in the form of \textit{Enthought Canopy}: a comprehensive Python analysis environment that provides easy installation of the core scientific analytic and scientific Python packages, creating a robust platform you can explore, develop, and visualize on. 

\subsection{The structure of the program}

One can divide the way the program works in three parts:
\begin{enumerate}
\item The preparation of a data taking session (figure~\ref{fig:configs}).
\item The actual data taking and partial processing (figure~\ref{fig:takedata}).
\item The calculation of the gain ratios and output of supplementary graphs (figure~\ref{fig:calculateratios}).
\end{enumerate}

\subsubsection{Preparation of a data taking session} 
The user can specify which part of the parameter space that the session should compromise. During one run the source is kept in a constant brightness range, also, the filterwheel positions are not changed. The only source parameters that is varied, is the \textit{Brightness DAC setting}. Furthermore one can select the gain configurations that should be included. The script creates a number of files which are used in step 2.

\subsubsection{Data taking and partial processing}

This is the part in which raw data is taken and processed to a format that can be used for higher level analysis. \textit{TakeData.py} first initialise some electronic components and resets all the sources. After this, it calls two modules in following order: \textit{GenerateData.py} and \textit{SummerizePD.py}. 

\begin{figure}
\includegraphics[width= \textwidth]{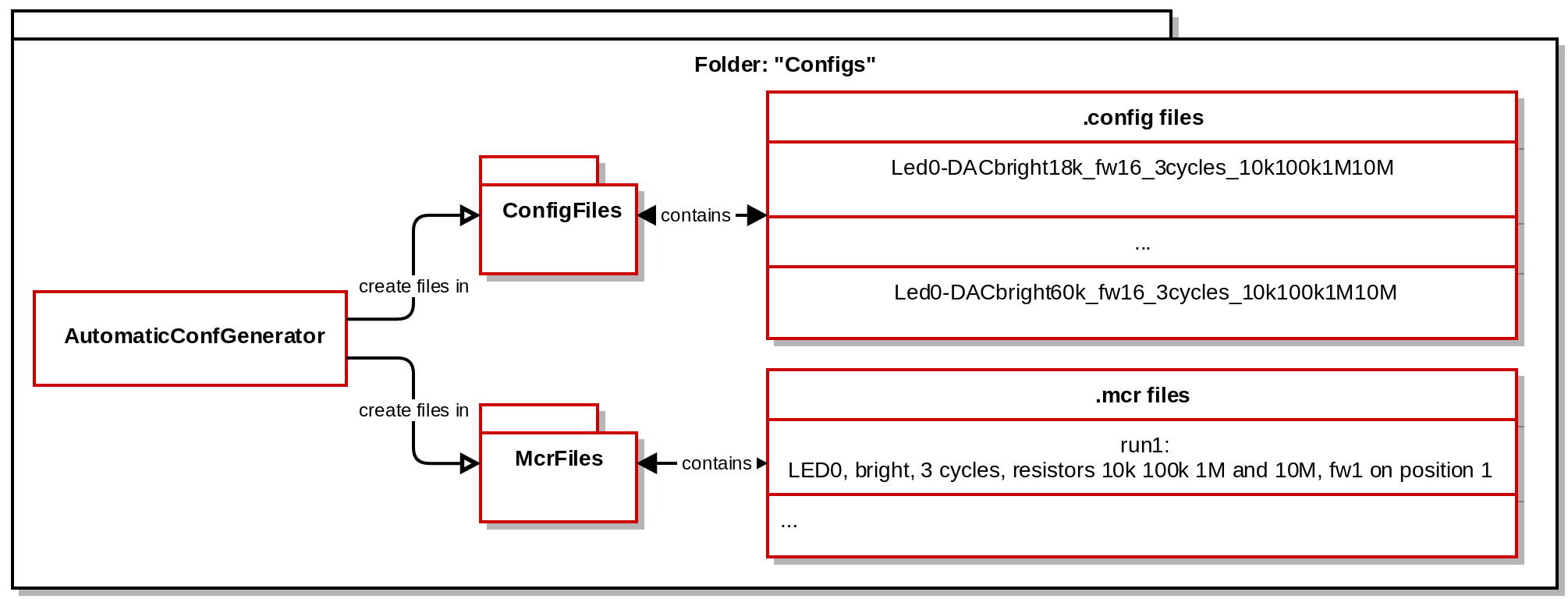}
\caption{Part one: the generation of configuration files that specify the parameters and will be used as input to start a data-taking session.}
\label{fig:configs}
\end{figure}

\begin{figure}
\includegraphics[width= \textwidth]{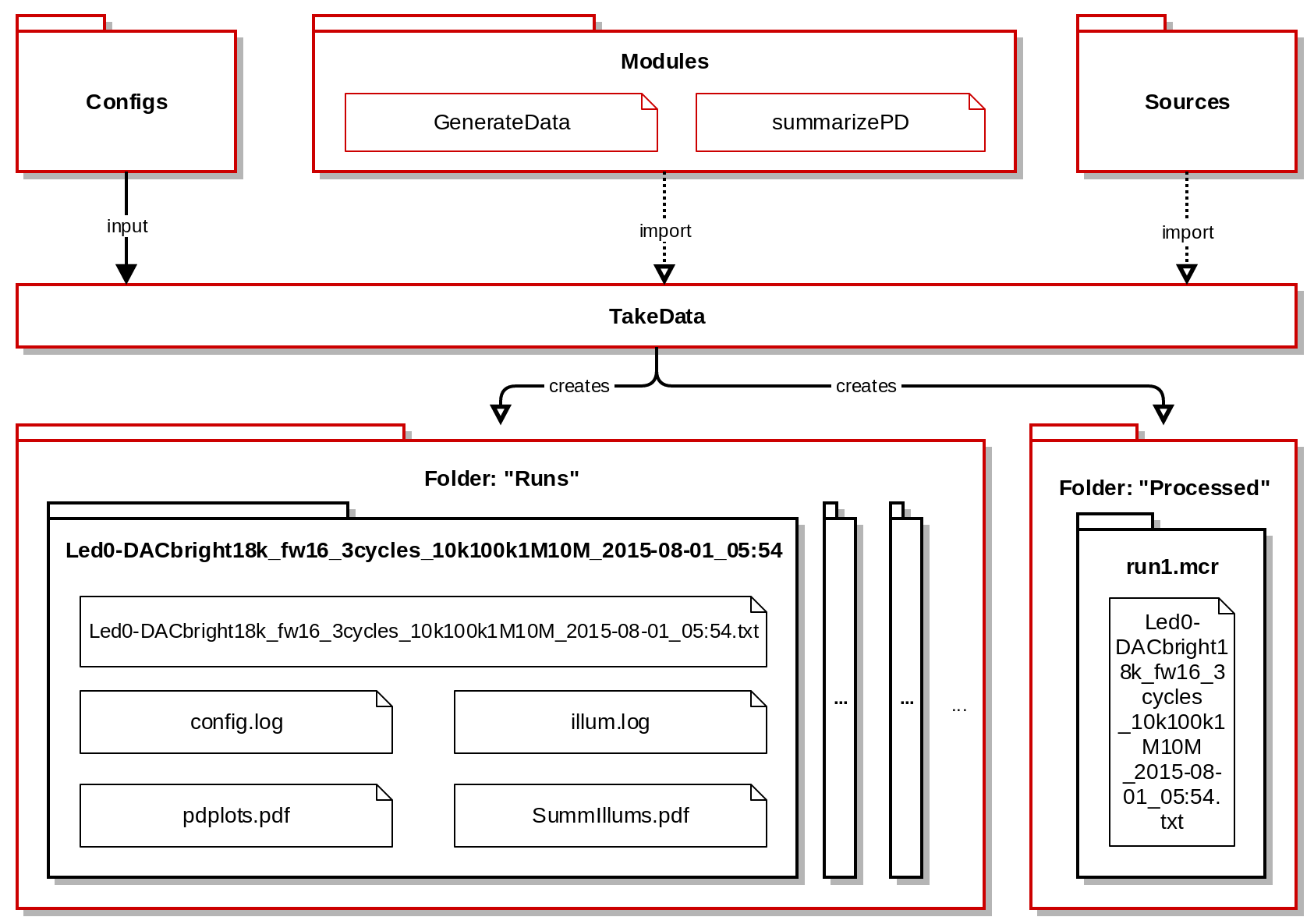}
\caption{Part two: \textit{TakeData.py} uses the files generated in the folder \textit{Configs} to do the actual measurements. It uses some lower level scripts for the handling of the readout hardware. The script produces a \textit{.txt}-file for each \textit{Brightness DAC setting} with a summary of the measurement. Alongside that, it also creates some files for diagnostic purposes.}
\label{fig:takedata}
\end{figure}

\begin{figure}
\includegraphics[width= \textwidth]{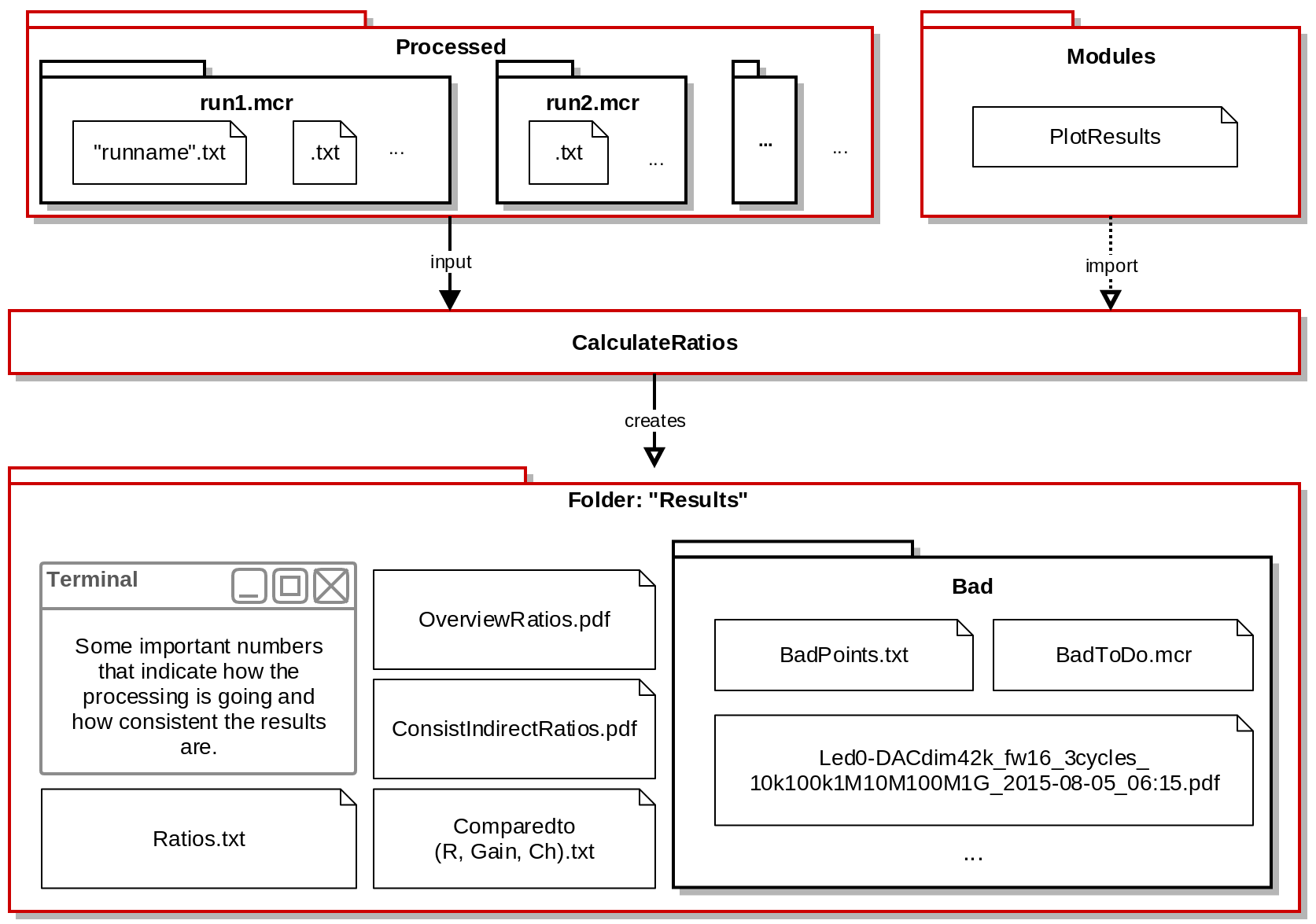}
\caption{Part three: \textit{CalculateRatios.py} processes the PD currents registered by \textit{TakeData.py} to ratios between gain configurations and the errors on those ratios. It is the part of the program where the actual analysis is done. The script has several higher level parameters to investigate the output and isolate possible causes of problems. To make this possible there is a rich variety of output available. }
\label{fig:calculateratios}
\end{figure}

\paragraph{\textit{GenerateData.py}} takes one \textit{.config}-file as input to specify the parameters. The light is turned on and off between all the measurements to determine the dark current. Besides that the source parameters are fixed. The total session can be divided in cycles. All gain configurations that are included in a cycle are specified in the input file (see figure~\ref{fig:configs}). Over those settings is looped in a random order. The randomization procedure has been introduced to investigate and reduce time dependent effects in later analysis. 
The raw output is once per second written to a file for each requested gain combination. It is formatted in two ways: the number of counts and the converted current value in fA. \textit{GenerateData.py} is included as an appendix.

\paragraph{\textit{SummerizePD.py}} is launched by \textit{TakeData.py} when \textit{GenerateData.py} has completed all the cycles. \textit{SummerizePD.py} can be regarded as a first step in the data processing chain. For each gain combination, it calculates an average of the difference between the \textit{off} and \textit{on} parts of a measurement block, see figure~\ref{fig:time}. Afterwards, this average is combined over different cycles. The resulting dark-subtracted average currents are reported for PD1 \textit{ch0} (x1) and \textit{ ch1} (x101) in fA. 
To accomplish this, several steps are needed~\footnote{Here explained for one gain configuration.}:
\begin{enumerate}

\item The averages of all the \textit{on} and all the \textit{off} blocks are calculated. The standard deviation in this group of points is calculated as $\sigma = \sqrt{\frac{Var}{N_{points}}}$. Where \textit{Var} is the variance of those points and $N$ is the number of points. In the default case, for both \textit{on} and \textit{off}, $N$ is 20.

\item If the \textit{on} block average was higher than 28000 counts, information for that configuration was not calculated further because of saturation effects.

\item The \textit{off} current was subtracted from the corresponding \textit{on} current. The new error used is simply $\sigma_{subtracted} = \sqrt{\sigma_{on}^2 + \sigma_{off}^2}$. If the resulting dark-subtracted average value is less than 50 counts, the configuration is skipped because the $signal/noise$ level is to low. 

\item The values that were found over different cycles are compared. There are several ways to combine these. This is currently implemented as follows:
\begin{enumerate}
\item Calculate the weighted mean of those \textit{block-averages}:
\begin{equation}
\mu = \frac{\sum_i \frac{x_i}{\sigma_i^2}}{\sum_j \frac{1}{\sigma_j^2}} \: \: \:  \: \: \: \text{and}  \: \: \:  \: \: \: \sigma^2_{\mu} = \frac{1}{\sum_i \frac{1}{\sigma_i^2}}
\end{equation}
\item Take the variance of those \textit{block-averages}, from this variance follows an extra spread 
\begin{equation}
\sigma_{cycles} = \sqrt{\frac{Var}{N_{cycles}}}.
\end{equation} 
\item The combined variance on $\mu$ is $\sigma_{stat} = \sqrt{\sigma_{cycles}^2 + \sigma_{\mu}^2}$.
\item At least, an additional error which takes the ADC count resolution into account has been added: $\sigma_{final} = \sqrt{\sigma_{stat}^2 + (1/2)^2}$.

\end{enumerate}
\item Convert the result to fA and write it to an output file. 
\end{enumerate}

\begin{figure}
\includegraphics[width= 0.7\textwidth]{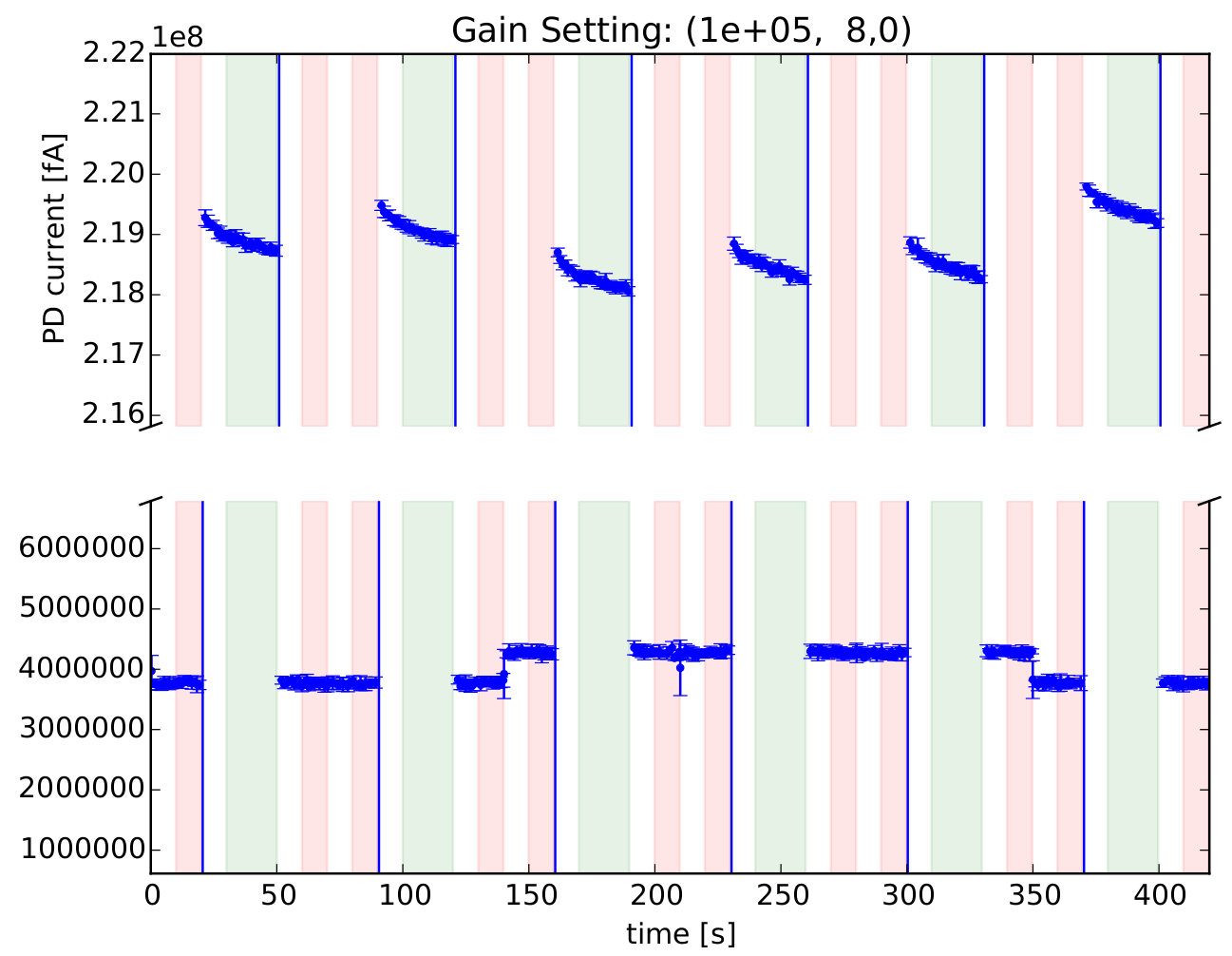}
\caption{PD1 currents for the gain setting (\textit{resistor gain} 100k, \textit{ADC gain} 8, \textit{ ch0 gain}). The source settings were Led0-DACbright20k\_fw16. There is a time gap between the double red \textit{off}-periods because multiple gain configurations were measured in between two cycles of the same settings. This measurements contains six cycles.}
\label{fig:saggering}
\end{figure}

The way the errors are calculated may seem a little heuristic. The current roadmap is proposed because more rigorous error estimates returned errors that were to small to be consistent. This could be explained by instabilities and slightly time dependant behaviour in the source section. \textit{SummerizePD.py} has several checks to monitor these unexpected non reproducibilities. 

There is check that looks for inconsistencies between the averaged PD current off-set before and after the pulse. The goal of this check is to eliminate possible drifts in the off-set. 

Another behaviour that is monitored and lead to the need for enlarged errors is the so called \textit{sagging} phenomenon, illustrated in figure~\ref{fig:saggering}. By \textit{sagging}, the drop in PD current during a pulse is meant. The amount of sagging in the \textit{on}-time region of a block is quantified by the $85^{th}$ percentile minus the $15^{th}$ percentile of the \textit{on}-time data. A more thoroughly discussion about this effect will will be postponed until chapter~\ref{sec:8}.

The final consistency check is between the resulting dark-subtracted averages of the different cycles. In the ideal case there should be no noticeable differences between two cycles of the same measurement. But, as can bee seen from figure~\ref{fig:saggering}, there were differences. Because we were unable to fully determine the origin of those differences they were included as an extra error. 

The program includes several options to make plots and register the worst events in different text files for further problem-solving purposes. 

\subsubsection{Calculation of the ratios between gain configurations}

As can be seen in figure~\ref{fig:calculateratios}, the input for the calculations of the ratios are the text files created in the \textit{processed} folder. For each file, \textit{CalculateRatios.py} calculates all the possible ratios with their error. The ratios are stored in the following way:

\begin{center}
\begin{tabular}{l|ccll}
 \rule{0pt}{3ex}       & highest total gain & lowest total gain & $\frac{I_{high}}{I_{low}}$ & $\sigma ( \frac{I_{high}}{I_{low}} )$ \\ \hline
example  \rule{0pt}{3ex} & (100k, 8, 1)       & (10M, 1, 0)         & 0.998                  & 0.041                         
\end{tabular}
\end{center}

Where (100k, 8, 1) represents the gain combination: resistor gain of 100k$\Omega$, read out channel 1 (x101) and ADC gain set to 8. $\frac{I_{high}}{I_{low}}$ represents the ratio of the dark-subtracted average currents of those two gain settings. Notice that there has been averaged over a lot of different source settings to obtain this ratio. Nevertheless, the ratio was assumed to be brightness independent so should give a consistent value for those different source configurations. 

Since there is carefully scant over the the parameter space of the source, there are a lot of measurements which differ only slightly in PD current. This enables us to have lots of values for most ratios. The weighted average and weighted error is calculated in the same way as was done previously. If a specific value of a ratio, coming from one fixed source configuration, is incompatible within three sigma deviations from the weighted mean, it is called a "Bad point" and indicated in red on most of the following graphs. Notice that those points are not thrown away for further analysis. All those bad points are grouped together and written to a file.

At the end of this procedure, a list of \textit{direct ratios} is found. The program first tries to isolate the causes of the \textit{bad points} in two ways. First of all it checks if a certain source setting gives rise to an unusual high number of \textit{bad points} in it. Secondly it searches for ratios that contain a relatively big amount of \textit{bad points}. Those source specific file names and those \textit{bad ratios} are also written to a file. There is also an option included to redo the measurements for the \textit{bad} source settings to see if peculiarities are reproducible. The thresholds for this selection were tuned during the project. Because there were some hints that indicated a possible problem with certain gain settings~\footnote{This will be discussed in section~\ref{sec:8}.}, the bad ratios calculated up to now were excluded from  the current list. 

The problem with the current list, certainly after the removal of \textit{bad ratios} is that it does not contain all the possible combinations, even compared to an intermediate reference gain configuration. another point of interest that needs to be addressed is that the errors grow fast when the difference between the lowest and the highest gain setting increases. This is due to the lack of source settings in the overlap region of both gain settings. Nevertheless, both problems can be greatly reduced by combining ratios to calculate new ratios that were out of reach or only roughly know before. As stated before, the total gain range is $\approx 8.08\pow{7}$, while the upper bound on the ratio of gains that can be calculated in a direct manner is $28000/50 \approx 500 $. This restriction is due to the limitations of the readout hardware. Taking an indirect way of calculating ratios into account could raise this towards an upper limit of $500^2 = 250000$. If we take the reference gain setting somewhere in the middle it should in theory be possible to compare that gain setting with all the other ones and determine the relative errors.

In practice this  is implemented as follows:
\begin{enumerate}
\item One gain is taken as reference gain, $G_R$, all the indirect ratios will only be calculated with respect towards this gain.
\item A list of the weighted averages of direct ratios that includes the reference setting is created.
\item Loop over al the gain combinations in this list. To clarify the procedure, lets focus on one gain setting on this list and call is $G_I$.
\item The total gain of this last one is compared towards the total gain of the one we are focussing on. Without loss of generality, we can assume that the last one is higher: $total\_gain(G_R) < total\_gain(G_I) $.
\item A list of the weighted averages of direct ratios that includes $G_I$ can now be constructed.
\item Only the settings with a higher gain than $G_I$ are considered in this list. Those higher setting under consideration will be called the final gain setting, $G_F$. We have: $total\_gain(G_R) < total\_gain(G_I) < total\_gain(G_F)$.
\item for each remaining choice of $G_F$ an \textit{intermediate ratio} with $G_R$ is calculated:
\begin{equation}
\frac{I_{G_F}}{I_{G_R}} = \frac{I_{G_I}}{I_{G_R}} \times \frac{I_{G_R}}{I_{G_I}}.
\end{equation}
The error on this ratio follows from standard rules of error propagation. Notice that the same process can also be used for gain settings with a lower total gain than $G_R$.
\item The double loop structure calculates all the possible combinations that can be reached by applying strategy.
\item A total list of \textit{indirect ratios} is constructed. Because in most occasions there are several ways~\footnote{Several intermediate gain settings, $G_I$, can be used to reach a final gain setting, $G_F$.} to reach a final setting $G_F$, a weighted mean and weighted error is calculated. 
\end{enumerate}

Notice that the condition $total\_gain(G_R) < total\_gain(G_I) < total\_gain(G_F)$ is not strictly necessary for the strategy. This is just an optimisation since leaving this one out would not lead to a larger number of gain combinations that could be calculated or to a higher precision on the ones that were already within reach. On the contrary, leaving out this condition increases the risk of deteriorating good direct gain calculations because of the combination with worse intermediate ones.

At this phase, there are two lists of ratios compared to the $G_R$ reference setting: the direct ones and the indirect ones. Those can be merged into one \textit{super combined list} by using the following set of rules:
\begin{itemize}
\item case 1: The gain setting under consideration is included in both lists. Both values are checked for compatibility, if they are compatible, the weighted average and weighted error is included in the \textit{super combined list}.
\item case 2: The gain setting only occurs in one of the lists. As it is not possible to check this value it is simply added to the \textit{super combined list}.
\end{itemize}

All lists mentioned are kept for further analysis. They are written to files in the folder \textit{Results}: \textit{Ratios.txt} and \textit{Comparedto(R,ADC-gain,Ch).txt} (see figure~\ref{fig:calculateratios}).

\section{\label{sec:7}Results}

Table~\ref{terminal} gives a quick overview of what could be expected from the results. It groups all the data of several weeks of data taking and only makes a hard distinction in the ADC hardware that was used. The full process was done with two different boards to further isolate the cause of some anomalies (see also section~\ref{sec:8}). 

The following parameters can be adjusted:
\begin{itemize}
\item \textit{sigmasingleratio = 3}: If we combine the error of an individual point with the error on the mean of a certain ratio, and the point lies further away than 3 times that combined error from the mean, this point is marked as red/bad point.
\item \textit{sigmacombinedratios = 1.5}: If a ratio can be calculated in both direct and indirect ways, those two weighted averages are checked for consistency before combining them. They are consistence if they lie in the interval of 1.5 times their combined error.
\item \textit{ratiobadpoints = 0.2}: Some ratios show strange behaviour. If 20\% of the direct measured points are not compatible, all the direct data of that ratio will not be used for further analysis.
\item \textit{numberofbadpointsinfile = 8}:	It is also suspicious if one fixed source settings run leads to a lot of bad data, this could mean that there are some problems with the led for example, files which lead to more than 8 bad point will be listed for analysis purposes.
\end{itemize}

\begin{table}
\centering
\caption{Some numbers that are used to fine tune the analysis parameters and serve as a quick way to see were things could be improved. The retrieved values for board 1 and board 2 are quite similar. A point is the division of two dark-subtracted, cycle-average0,d currents of different gain settings for fixed source parameters.}
\label{terminal}
\begin{tabular}{l|c|c}
                            \rule{0pt}{3ex}                     & ADC board 1 & ADC board 2 \\ \hline
Number direct ratio points        \rule{0pt}{3ex}               & 4129        & 3847        \\
Direct ratios plotted                          & 160         & 160         \\
Incompatible direct ratio points (bad points)  & 503         & 599         \\
Ratios with only one direct ratio point        & 3           & 0           \\
Unique ratios with more than 20\% bad points   & 25          & 27          \\
Incompatible direct ratio points in bad ratios & 418/503     & 498/599     \\
Leftover direct ratios (kept for combining)   \rule{0pt}{3ex}   & 132         & 133         \\ \hline
Direct gain settings compared with $G_R$     \rule{0pt}{3ex}     & 14          & 15          \\
Number of intermediate ratios for $G_R$         & 67          & 71          \\
Indirect gain settings compared with $G_R$      & 20          & 21          \\
\end{tabular}
\end{table}

\subsection{Overview of the direct ratios}

\begin{figure}
\includegraphics[width= 0.7\textwidth]{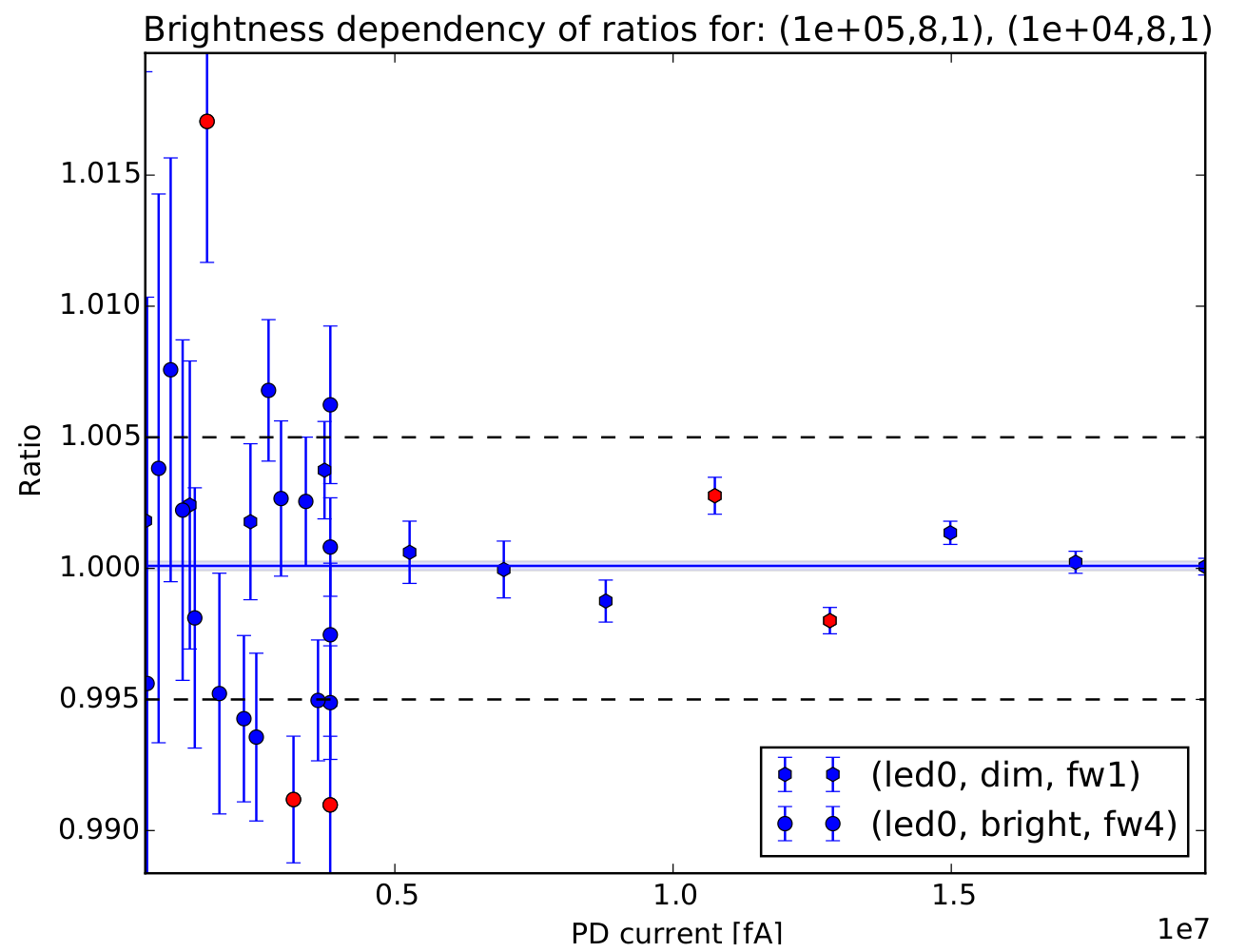}
\caption{Example of one of the 160 ratios that could be calculated in a direct manner. The figure uses data for the two different gain settings indicated in the title and evaluates their ratio in a lot of different source conditions: filterwheel position, LED brightness range and LED brightness DAC setting are varied. The dotted lines indicate the 0.5\% error zone to guide the eye. The blue line is the weighted average of the points, including the red ones, and the small, more shallow, blue zone around it is the weighted average error on the mean. \textit{Bad points} are indicated in red.}
\label{fig:over2_good}
\end{figure}

One of the most important output files of \textit{CalculateRatios.py} is \textit{OverviewRatios.pdf}. It gives a visual overview of all the direct ratio combinations and is a good tool to spot unexpected behaviour. An example can be seen in figure~\ref{fig:over2_good}.

\subsection{Consistency of indirect ratio calculations}

\begin{figure}
\includegraphics[width= 0.7\textwidth]{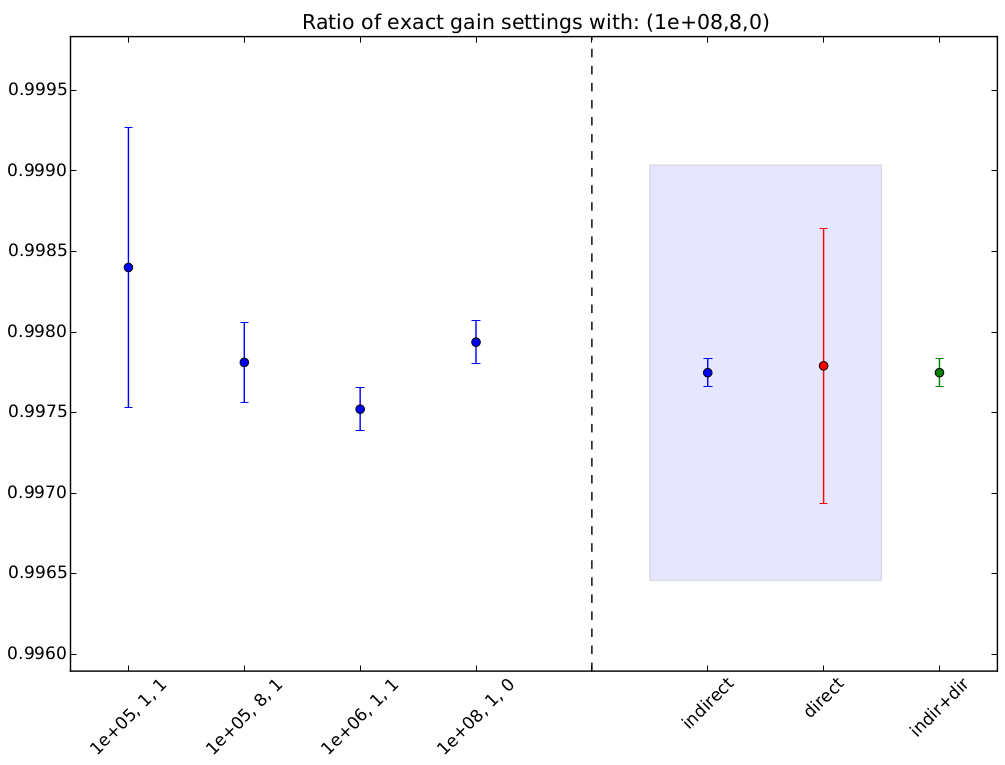}
\caption{Example of one of the ratios that could be calculated in both direct and an indirect manner. This kind of plots enables a more graphical approach to check the consistency. On the right side the indirect calculations are shown separately with their error. In this case there were four possible gain configurations that could be used as intermediate gain. On the left side, the weighted average is given and compared with the direct measurement. The shaded box indicates 1.5 times the combined error. In this cases the two are compatible and are combined.}
\label{fig:combi2_good}
\end{figure}

Since the different intermediate ways to calculate a gain were not compared with each other automatic before combining them to a weighted indirect value, it was important to have a graphical representation of this process. An example is showed in figure~\ref{fig:combi2_good}. 

\subsection{Summary of the final output}

\begin{figure}
\includegraphics[width= 0.7\textwidth]{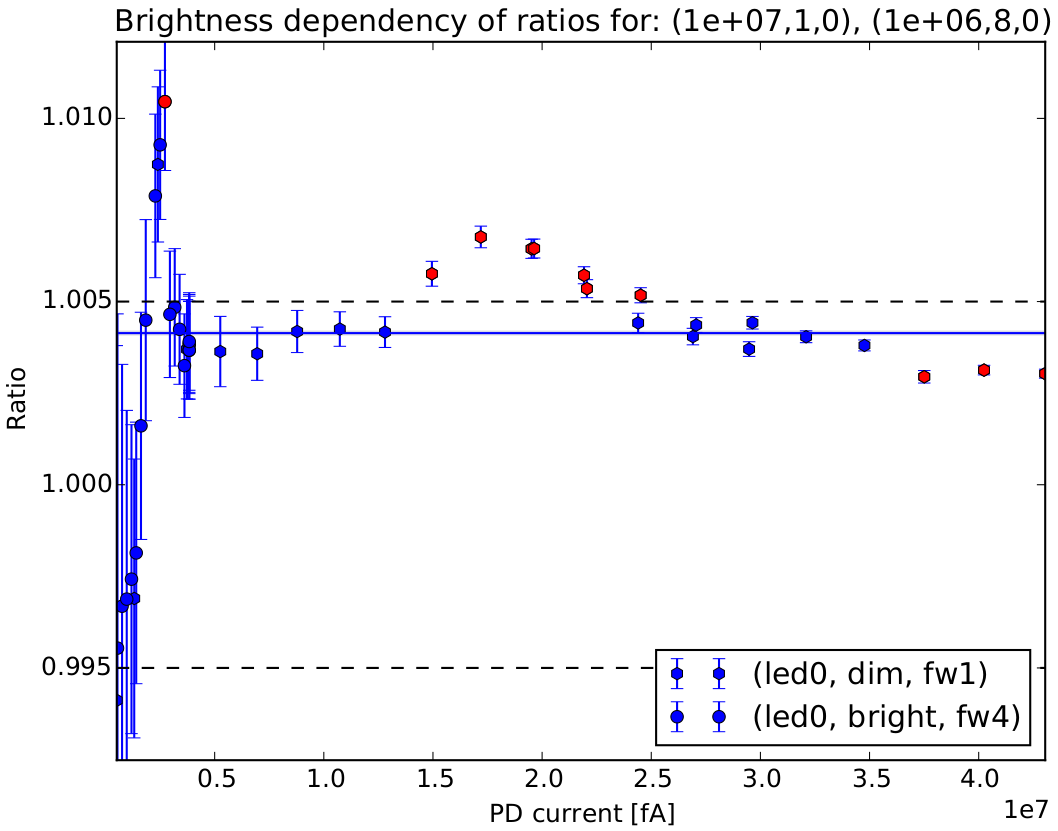}
\caption{The direct ratio graph corresponding with the empty line of gain setting \textit{(1M,8,0)} in table~\ref{supercombinedtable}. Because there are more than 20\% of the point inconsistent when compared with the weighted average, all the data in this plot was automatically removed from the indirect ratio calculations.}
\label{fig:over2_res}
\end{figure}

If the built-in checks at all stages of the program are reassuring and a the produced graphs are evaluated for peculiarities, the final textual output could be trusted. After one has designated one gain setting the status of reference gain, ratios towards the other settings are calculated. In table~\ref{supercombinedtable} the output is given for ADC board 2 and reference gain setting \textit{(10M,1,0)}. Notice that one of the empty rows is the reference gain, this is a trivial row of course. The reason that the ratio with respect to the setting \textit{(1M,8,0)} could not be calculated is more concerning. This is the setting that has a lower total gain value as close as possible to the reference one. For that reason it could not be calculated indirectly. This leads to the fact that this ratio was skipped in the final analysis because more than 20\% of the points that were used were incompatible with the weighted average. The overview of those points for the ratio under consideration is showed in figure~\ref{fig:over2_res}. A more careful analysis of this problem is postponed to the next section. 

\begin{table}
\begin{center}
\caption{Information contained in \textit{Comparedto(10M,1,0).txt}. The reference gain setting used was: resistor gain $10\pow{6} \Omega$, ADC gain x1 and channel 0 (x1). Of the 24 combinations that were tested, $G_R$ could be compared with 21 others, only the highest gain setting $(1G, 8, 1)$ was out of reach.}
\label{supercombinedtable}
\begin{tabular}{ccccccc}
\hline
	& \multicolumn{2}{c}{\bf{Direct}} & \multicolumn{2}{c}{\bf{Indirect}} & \multicolumn{2}{c}{\bf{Combined}} \\
Gain & ratio & $\sigma$ & ratio & $\sigma$ & ratio & $\sigma$ \\
\hline
(10k, 1, 0)	&		/	&			/		& 0.99869  & 0.00056		& 0.99869 &  	0.00056  \\
(10k, 8, 0)   &   1.00404  & 	0.00666		&  0.99547  & 0.00021	& 0.99548  & 	0.00021 \\
(10k, 8, 1)  &     1.00005 &     0.00008		& 		/	&		/	& 1.00005 &  	0.00008 \\
(10k, 1, 1)   &   0.99990   &	0.00030 		& 1.00015  & 	0.00011 & 1.00012  & 	0.00010 \\

(100k, 1, 0)  &   0.99958   &	0.00370 		& 0.99893  & 	0.00017 & 0.99893 &  	0.00017\\
(100k, 8, 0)  &   0.99532   &	0.00033  	& 0.99400  & 	0.00012 &  0.99532  & 	0.00033\\
(100k, 1, 1) &    0.99985  & 	0.00004 		&	/		&		/	& 0.99985 & 	0.00004\\
(100k, 8, 1)  &   0.99975   &	0.00008  	& 0.99979 &  	0.00009 & 0.99977 &  	0.00006\\

(1M, 1, 0)   &    0.99973   &	0.00026 		& 0.99997 &  	0.00010 & 0.99994  & 	0.00009   \\
(1M, 8, 0)  &		/	&			/		&		/	&		/	&/&/\\
(1M, 1, 1)    &   0.99931   &	0.00009   	& 0.99931  & 	0.00010  & 0.99931 &  	0.00007\\
(1M, 8, 1)    &   0.99954   &	0.00086 		& 1.00076  & 	0.00008 & 1.00075  & 	0.00008   \\

(10M, 1, 0)  &		/	&			/		&		/	&		/	&/&/\\
(10M, 8, 0)  &    0.99891   &	0.00007  	& 0.99896  & 	0.00008 & 0.99893 &  	0.00005  \\
(10M, 1, 1)  &	 0.99967 &	    0.00099   	& 1.00068 &  	0.00008 & 1.00067  & 	0.00008 \\
(10M, 8, 1)  &		/	&			/		& 0.99892   	&   0.00047  & 0.99892  & 	0.00047 \\

(100M, 1, 0)  &   0.99905   &	0.00009 		& 0.99901 &  	0.00006  & 0.99903   &	0.00005   \\  
(100M, 8, 0) &    0.99779  & 	0.00086   	& 0.99775 &  	0.00009  &  0.99775   &	0.00009   \\
(100M, 1, 1)  &		/	&			/		& 0.99780 &  	0.00042 &  0.99780   &	0.00042  \\
(100M, 8, 1)  &		/	&			/		& 0.99147  & 	0.00062 & 0.99147 &  	0.00062   \\ 
 
(1G, 1, 0)   &    1.00005  & 	0.00099   	& 1.00107  & 	0.00008  & 1.00106  &	0.00008 \\
(1G, 8, 0)  &		/	&			/		& 0.99777  & 	0.00047 & 0.99777  & 	0.00047\\
(1G, 1, 1)  &		/	&			/		& 0.99137  & 	0.00064 &	0.99137   &	0.00064\\
(1G, 8, 1)  &		/	&		/			&  		/	&		/	&/&/\\
\hline
\end{tabular}
\end{center}
\end{table}

\section{\label{sec:8}Analysis of problems}

At first sight the textual results of the program look very promising. In most cases this is indeed believed to be the case, but there were some hints that not everything was consistent and that the set-up had some peculiar behaviour when pushed to the limits. A large proportion of time was invested in the development of methods to identify, and where possible, eliminate these peculiarities. Despite the effort not all problems could be completely resolved. The three most important ones will be disclosed in this section.

\subsection{The sagging of LED output during ON periods}

\begin{figure}
\includegraphics[width= \textwidth]{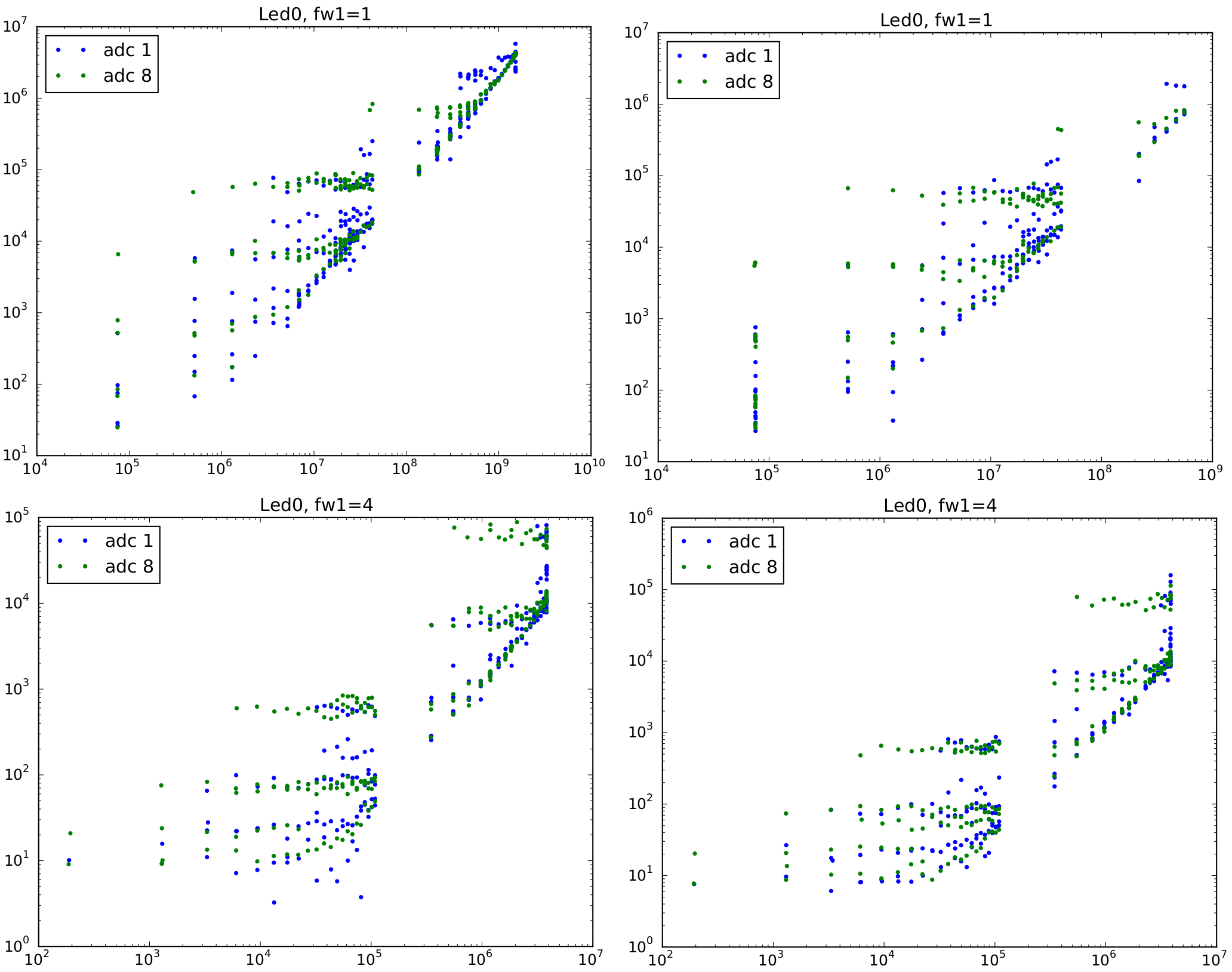}
\caption{The absolute amount of sagging, as defined in the text, plotted in function of PD1 current. The panels on the left show data taken with the first board while the data on the right was taken with the second one. A distinct plot is made for the two different filterwheel positions. Colours indicate the ADC gain settings used for the particular data points.}
\label{fig:sagging}
\end{figure}

The sagging phenomenon was already briefly discussed in the context of figure~\ref{fig:saggering}. \textit{Sagging} refers to the drop in PD current during a pulse. The minimal amount of sagging in the \textit{on}-time region of a block is quantified by the $85^{th}$ percentile minus the $15^{th}$ percentile of the \textit{on}-time data. This has to be averaged over the cycles:
\begin{equation}
\frac{\sum_{cycles} (p_{85}(I_{on})-p_{15}(I_{on}))}{N_{cycles}}.
\end{equation}
This quantity is plotted in figure~\ref{fig:sagging}. The amount of sagging of the PD current in each ON period is linearly correlated to the brightness and does, according to this first analysis, not depend on the gain settings. For each brightness one can put a lower limit on the amount of sagging that will occur. The independence of the ADC board used for the analysis is striking. This does highly suggest that the problem is caused by instabilities in the source sector. This should be tested in the future by repeating the analysis with LED 3 or the laser. Beside this, a structure of many branches appears towards larger sagging, this has yet to be explained by a more thorough study.

\subsection{The anomaly at low counts in the low brightness range}

\begin{figure}
\includegraphics[width= 0.7\textwidth]{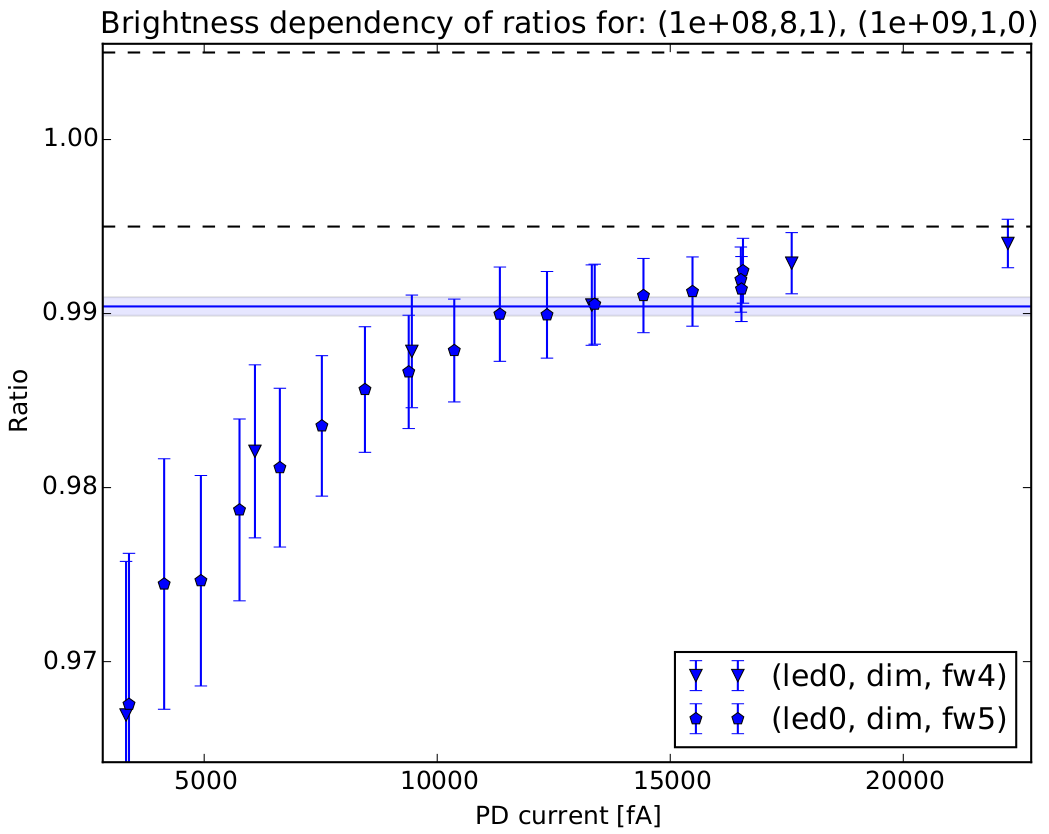}
\caption{An example of the anomaly at low counts in the low brightness range. The effect is independent of the filterwheel position and the DAC brightness setting of the source.}
\label{fig:tail}
\end{figure}

Bellow a PD current of approximately 20000 fA the ratio tends to drops. This means that the lower gain setting has too much counts or the higher setting too few. The current suggestion is an anomaly of the hardware in cases where the gain settings used are relatively high in combination with measurements where the number of counts is relatively low. In this situation the count rate for the lowest gain setting is thought to be slightly overestimated, resulting in an error on the gain ratio up to 4\%. The effect appears with both ADC board 1 and ADC board 2. An example of the situation is depicted in figure~\ref{fig:tail}. The ratios where this effect showed up and those where it does not are listed in the manual. This division in two groups is completely identical for both ADC boards.

The importance of this peculiarity arises because of the long term goal of the set-up: calibrating the actual DOMs in a more precise way than ever done before. For this purpose, the most dim source regions and the highest gain settings are indispensable.

\subsection{The ADCx8 Channel 0 problem}

\begin{figure}
\includegraphics[width= 0.7\textwidth]{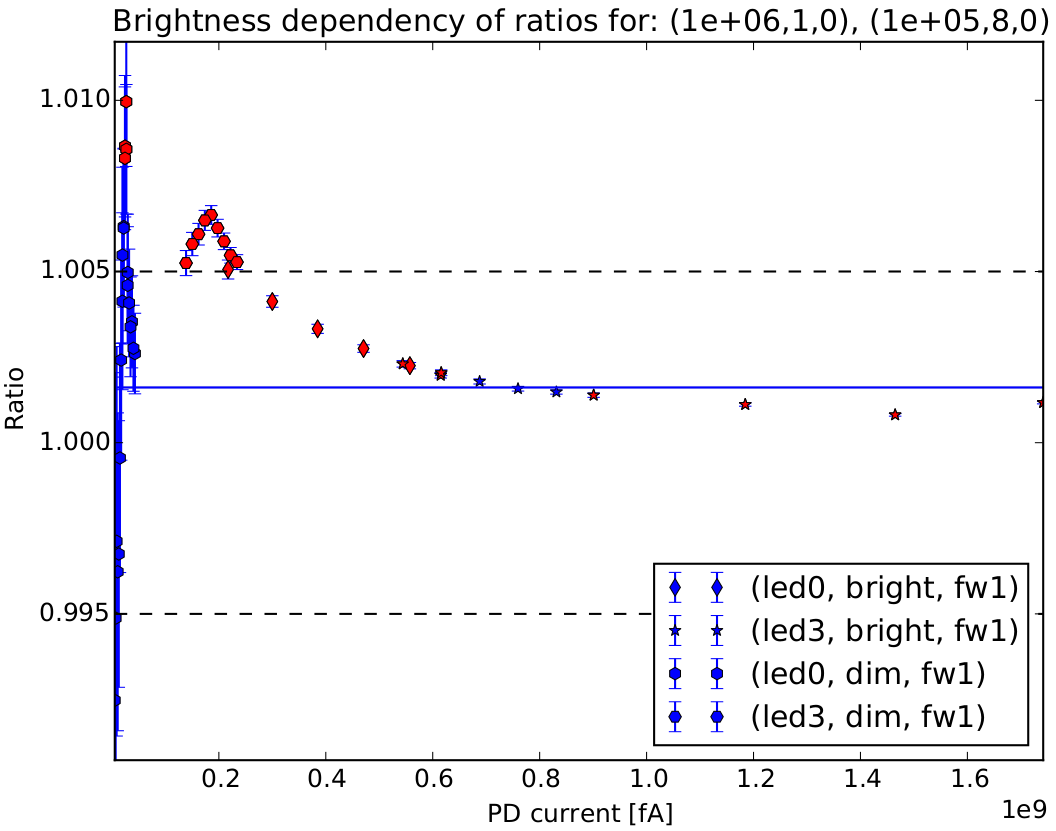}
\caption{An example of the ADCx8 Channel 0 problem. From this example, one can immediately see that the problem is independent of the LED that was used.}
\label{fig:adc8}
\end{figure}

A lot of bad ratios have something to do with the combination of ADCx8 and Ch0. In a certain brightness range, dependent on the resistor gain, this gives strange behaviour. The range corresponds to a fixed number of ADC counts around 2000-8000. So $brightness \times total\_gain$ is a constant for this problem. By looking for overlap in this zone for different source settings we were able to exclude the source as a cause of it.
The problem is reproducible with different LEDs, filterwheel positions and even ADC boards. An example is given in figure~\ref{fig:adc8}. Notice that the magnitude of the deviations are around 0.5\% for these anomalies. Currently this effect is thought to be caused by non-linearities in the amplifying circuit.

\section{\label{sec:9}Conclusion and outlook}

A laboratory set-up has been developed to more precisely measure the DOM optical sensitivity as a function of angle and wavelength. DOMs are calibrated in water using a broad beam of light whose intensity is measured with a NIST calibrated photodiode. This study will refine the current knowledge of the IceCube response and lay a foundation for future precision upgrades to the detector. Good understanding of PD readout is indispensable for DOM calibration. The main goal of the project was to investigate corrections on the photodiode measurements due to the amplifier circuit. To accomplish this, a general software structure has been added to the already existing framework of the  laboratory set-up. Since the set of parameters in the source sector is still growing, modularity and a high level of automation were important objectives. The software features a large array of graphical tools to intercept problems at a low level while the analysis can be easily adapted to the needs of foreseeable situations. A manual has been written that will guide the further development of the PD software. 

At current stage the errors in the different amplification chains are almost everywhere determined up to the level of 0.1\%. The brightness averaged ratios of the different gain configurations are almost always within a 1\%  deviation, in accordance with expectations and demands for the experiment. 

Nevertheless some ratios showed unexpected brightness dependence. The peculiarities could be isolated but not fully resolved. Further studies should be able to diminish these effects by making hardware improvements to both the source as the amplification sector. Already with the current outcome, it is perfectly possible to select certain gain configurations and brightness levels for the final calibration of the optical IceCube modules. Those settings can be trusted over a large range of brightnesses with  errors less than 0.1\% on their ratio compared with the \textit{ideal} gain setting. The ratios found in this study are then implemented in the overall calculations concerning the DOM calibration. This study has carefully documented those settings and their corresponding brightness range.

\clearpage
\appendix
\section{\label{sec:A}Greissen-Zatsepin-Kuzmin (GZK) cut-off}

\begin{figure}
\includegraphics[width=0.6\textwidth]{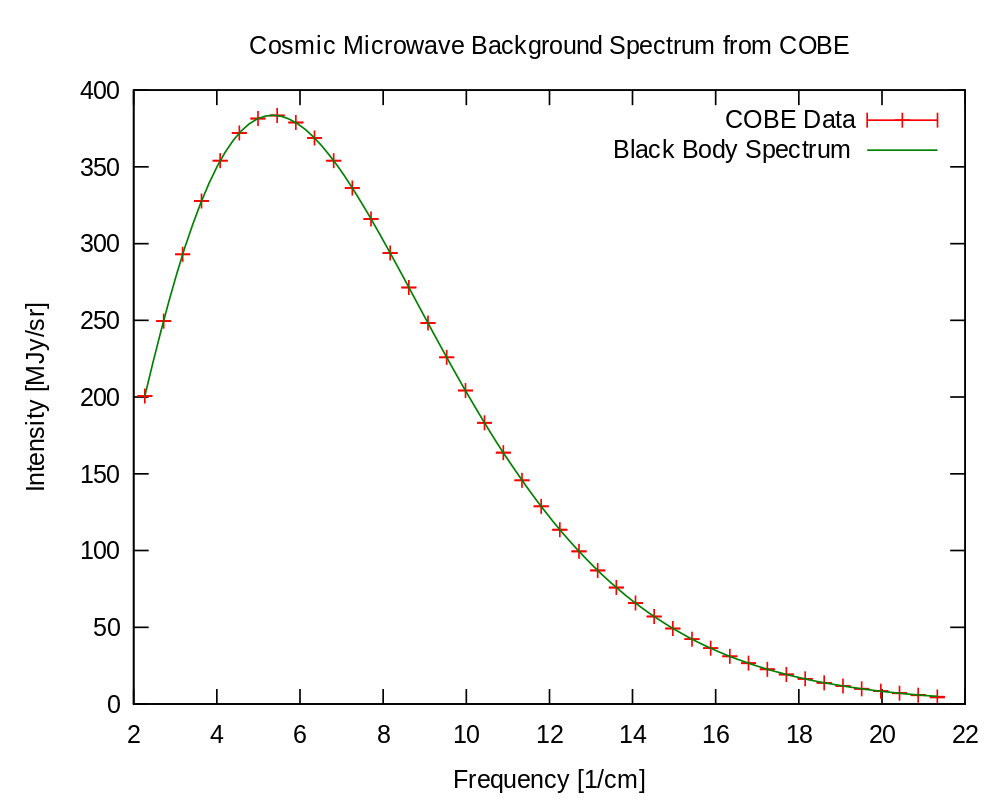}
\caption{Graph of the cosmic microwave background spectrum, measured by the FIRAS instrument on the COBE. The spectrum can be perfectly fitted to a black body spectrum at a temperature of around 2.7K.}
\label{fig:blackbody}
\end{figure}

Consider a high energetic proton as a cosmic ray particle. As space is permeated with cosmic background radiation, we could expect that the proton will scatter off such a photon. Throughout the calculation we will assume that the initial energy of the proton is high enough to treat it purely relativistic, we will check this afterwards. During a scattering process, new particles can be created without violating conservation of energy and momentum. As we are interested in high energy neutrinos, consider the following reaction:
\begin{equation}
p^+_{CR}+\gamma_{CMB} \rightarrow \Delta^+ \rightarrow n^0 + \pi^+ \rightarrow n + \mu^+ + \nu_\mu
\end{equation}
The created neutron is free and will decay to $p^+ + e^- + \bar{\nu_e}$ in almost al cases, the mean lifetime of this decay is $881.5 \pm 1.5$ s. The muon will decay as: $\mu^+ \rightarrow e^+ + \nu_e + \bar{\nu_\mu}$. The neutrino of interest for the IceCube observatory is the first one as this one has the chance to be very energetic if the initial proton was very energetic. So, the question of interest is: what is the minimum proton energy that is required to let this reaction take place?
As this is only an estimate we will use a fixed energy of 3 K for the CMB photon. In reality this energy follows a black body distribution with a peak around 2.725 K, this fit is illustrated in figure~\ref{fig:blackbody}. Conservation of relativistic momentum in the centre of mass leads to:
\begin{align}
(p_p+p_\gamma)^2 &= (p_n + p_\pi^+)^2 \\
m_p^2 + 2 p_p \cdot p_\gamma &= (m_n+m_\pi)^2 
\end{align}
The right hand side of the equation simplified because the lowest proton energy that can yield these two particles will produce them both at rest in the centre of mass frame. And as the relativistic momentum is Lorentz invariant, we can choose the frame in which we want to do the calculations.
To maximise the energy available from the collision, we make the momenta of the two particles in opposite directions. The 4–momentum of the proton is $(E_p,E_p)$ and of the photon $(E_\gamma,-E_\gamma)$
\begin{align}
2 p_p \cdot p_\gamma &= (m_n+m_\pi)^2 - m_p^2 \\
4 E_p E_\gamma &= (m_n+m_\pi)^2 - m_p^2
\end{align}
Which finally leads to:
\begin{equation}
E_p > \frac{(m_n+m_\pi)^2 - m_p^2}{4E_\gamma}
\end{equation}
\begin{itemize}
\item $E_\gamma = 3 \text{K} =  2.63\pow{-10}$MeV
\item $m_p=938.27$ MeV
\item $m_{\pi^+}= 139.57$ MeV
\item $m_n = 939.57$ MeV
\end{itemize}
This gives an estimate for the lower bound proton energy: $\approx 3\pow{20}$ eV.
As anticipated,this value is only a rough estimate, the statistical spread of the photon energy, especially the large tails towards smaller wavelengths, means that there are CMB photons with a much higher energy. This will lower the bound on the proton energy. Also, the process that we have
discussed, $ p + \gamma \rightarrow \pi^+ + n$, is not the only microwave background scattering process for high energy protons. In particular, a second process, $ p + \gamma \rightarrow p + \pi^0$, also takes place, and is energetically preferred because the final state particles are lighter. Taking into account these and other details, the energy at which you begin to see suppression of GZK photons is in fact around $3\pow{19}$ eV.

One can now proceed to calculate the mean propagation length that such a proton would be able to travel in space. For this, one would need an estimate of the cross section for pion production. The Breit-Wigner formula with the lowest lying nucleon resonance $\Delta^+ $ as intermediate state can be used. The final result is in the order of ten megaparsec. This leads us to a tentative conclusion that we cannot use UHE cosmic rays to look for extreme sources beyond our own cluster.

\section{\label{sec:B}Operational Amplifiers}

\begin{figure}
\includegraphics[width=0.45\textwidth]{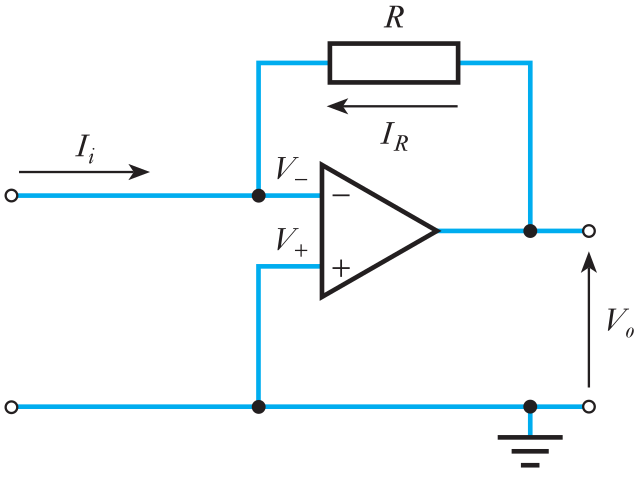}
\caption{A current-to-voltage converter~\cite{electronics}.}
\label{fig:opamp}
\end{figure}

operational amplifiers (or op-amps) are among the most widely used building blocks for the construction of electronic circuits. A comprehensive introduction to the various facets of these integrated circuits can be found in any basic electronics textbook~\cite{electronics}. A specific type of op-amp that should be mentioned in this work is the current-to-voltage converter. Some sensors, like photodiodes for example, operate such that the physical quantity being measured is represented by the magnitude of the current produced at its output, rather than by the magnitude of a voltage. This illustrates one of many situations where we may wish to convert a varying current into a corresponding varying voltage. A circuit to perform this transformation is shown in figure~\ref{fig:opamp}. The input current can be related to the output voltage in the case of an idealised op-amp:
\begin{equation*}
V_0 = -I_iR
\end{equation*}
Thus, the output voltage is directly proportional to the input current and the gain is proportional to the resistance used.

\section{\label{sec:C}Code excerpts and manual}
The tree of python files in the project directory on the local hub computer takes the following form:
\begin{itemize}
\item Configs
\begin{itemize}
\item \textit{AutomaticConfGenerator.py}
\end{itemize}
\item Modules 
\begin{itemize}
\item \textit{GenerateData.py}
\item \textit{summarizePD.py}
\item \textit{PlotResults.py}
\end{itemize}
\item \textit{TakeData.py}
\item \textit{CalculateRatios.py}
\item SpecialCases
\begin{itemize}
\item \textit{summarizePD\_extraplot.py}
\item \textit{Plot\_RawData.py}
\item \textit{Plot\_RawData\_OneGain.py}
\item \textit{Datadivider.py}
\item \textit{SoftwareVersion.py}
\item \textit{callSummPD.py}
\end{itemize}
\end{itemize}
Also the more low level file \textit{Sources.py} was adapted to make later applications and extensions more practical.
The python code of two important python parts of the program are given in this appendix: \textit{GenerateData.py} and \textit{CalculateRatios.py}.

To support further development of the program, an extensive manual was written with guidelines and technical details of the code. Both the manual, the full data and the code are available on request.



\section*{Supplementary material}

**GenerateData.py**

```python
  1: #!/usr/bin/python
  2:
  3: from time import gmtime, strftime
  4: from Modules import summarizePD
  5: import os,time,sys,string
  6: from random import shuffle
  7: from copy import deepcopy
  8:
  9: def getGlobalChannel(pd, ch):
 10:     if pd ==1:
 11:         globalChannel = ch
 12:     elif pd ==2:
 13:         globalChannel = ch+2
 14:     elif pd ==3:
 15:         globalChannel = ch+4
 16:     return globalChannel
 17:
 18: def convertPreampconfig(a):
 19:         b=[]
 20:         for i in range(0,len(a)):
 21:                 b.append(("B",a[i]))
 22:         return b
 23:
 24: def logData(source_state, count, pd,fAPerCount,d,t0):
 25:         nch=2
 26:         while count>0:
 27:                 start_time=time.time()
 28:                 sys.stdout.write("%f\t%d"%(start_time-t0, source_state))
 29:                 avgList=d.getADCAveragesReset().split(" ")
 30:                 #print "\nreadout: ", avgList , "\n"
 31:                 for ch in range(nch):
 32:                         globalCh = getGlobalChannel(pd, ch)
 33:                         avg=float(avgList[globalCh*3+1])*fAPerCount[pd, ch]
 34:                         sys.stdout.write(string.ljust("\t%.8g"%avg,9))
 35:                         sigma=float(avgList[globalCh*3+2])*fAPerCount[pd, ch]
 36:                         sys.stdout.write(string.ljust("\t%.8g"%sigma,9))
 37:                 for ch in range(nch):
 38:                         globalCh = getGlobalChannel(pd, ch)
 39:                         #   print "\nglobal channel for pd %i and ch%i is %i" %(pd, ch, globalCh)
 40:                         avg=float(avgList[globalCh*3+1])
 41:                         sys.stdout.write(string.ljust("\t%.8g"%avg,9))
 42:                         sigma=float(avgList[globalCh*3+2])
 43:                         sys.stdout.write(string.ljust("\t%.8g"%sigma,9))
 44:                 sys.stdout.write("\n")
 45:                 sys.stdout.flush()
 46:                 delay=start_time+1-time.time()
 47:                 if delay>0: time.sleep(delay)
 48:                 count-=1
 49:
 50: def setupConfigs(preampConfigs, pd, adcGain,d):
 51:         nch=2
 52:         fAPerCount = {}
 53:         preampType,transimpedance=preampConfigs
 54:         d.setPreampType(pd-1,preampType)
 55:         d.setTransimpedance(pd-1,transimpedance)
 56:         for ch in range(nch):
 57:                 globalCh = getGlobalChannel(pd,ch)
 58:                 d.setADCGain(globalCh,adcGain)
 59:                 fAPerCount[pd,ch] = d.fAPerCount(globalCh)
 60:         return fAPerCount
 61:
 62: def WriteConfigFromDict(pd,date,startTemp,endTemp,**dict):
 63:         string = "comment\t%s"%dict['RunName'] + "\n"
 64:         string+= "config\tstart\t%s"%str(date) + "\n"
 65:         string+= "config\tledNumber\t%i" % dict['led'] + "\n"
 66:         string+= "config\tbrightnessrange\t%s" % dict['brightnessrange'] + "\n"
 67:         string+= "config\tbrightness\t%i" % dict['brightness'] + "\n"
```

**GenerateData.py**

```
 68:            string+= "config\tfw1position\t%i" % dict['fw1position']+"\n"
 69:            string+= "config\tfw2position\t%i" % dict['fw2position']+"\n"
 70:            string+= "config\tsettling_time\t%d"%dict["settling_time"]+"\n"
 71:            string+= "config\ton_time\t%d"%dict["on_time"]+"\n"
 72:            string+= "config\toff_time\t%d"%dict["off_time"]+"\n"
 73:            string+= "config\tncycles\t%i"%dict["ncycles"]+"\n"
 74:            string+= "config\tstartTemp\t%d\tendTemp\t%d\n"%(startTemp,endTemp)
 75:            string+= "config\tscanAxes\t%s\t%s"%("rG","adcGain")+"\n"
 76:            string+= "config\tscanNPoints\t%d\t%d"%(len(dict['preampConfigs']),len(dict['adcGains']))+"\n"
 77:            string+= "config\tPD\t%i"%pd+"\n"
 78:            return string
 79:
 80:  def generatedata(d,m,source,**dict):
 81:
 82:            startTemp=d.getPreampTemp(0)
 83:            pd=2                           #number of adcboard
 84:            RunName=dict['RunName']
 85:            settling_time = dict["settling_time"]
 86:            off_time=dict["off_time"]
 87:            on_time=dict["on_time"]
 88:
 89:            fAPerCount ={}
 90:            preampConfigs = {}
 91:            os.mkdir(os.path.join("Runs",RunName))
 92:
 93:            sourceParameters={"lednumber":dict['led'], "brightnessrange":dict['brightnessrange'], "brightness":dict['brightness'],"fw1position":dict['fw1position'], "fw2position":dict['fw2position']}
 94:            source.setParameters(**sourceParameters)
 95:            print "source:\t%s"%'\t'.join(str(s) for s in source.getParameters())
 96:
 97:            #Make illum.log file
 98:            illum=open(os.path.join("Runs",RunName,'illum.log'),'w')
 99:            oldstdout = sys.stdout
100:
101:            #Make config file
102:            date=strftime("%Y-%m-%d %H:%M:%S", gmtime())
103:            preampConfigs=convertPreampconfig(dict['preampConfigs'])
104:
105:            for i in range(0,dict["ncycles"]):
106:                    tempreslist=deepcopy(dict['preampConfigs'])
107:                    shuffle(tempreslist)
108:                    for R in tempreslist:
109:                            tempadclist=deepcopy(dict['adcGains'])
110:                            shuffle(tempadclist)
111:                            for x in tempadclist:
112:                                    fAPerCount[dict['preampConfigs'].index(R)]=setupConfigs(preampConfigs[dict['preampConfigs'].index(R)],pd,x,d)
113:                                    sys.stdout=oldstdout
114:                                    print "rg%s_adc(ch0)%s_adc(ch1)%s:\t cycle %i of %i" %(R,x,x,i+1,dict["ncycles"])
115:                                    illum.write("start\t%s\t%i%i\t%i\t%i"%(R,x,x,dict['preampConfigs'].index(R),dict['adcGains'].index(x)))
116:                                    illum.write("\n")
117:                                    t0=time.time()
118:
119:                                    sys.stdout=illum
120:                                    # Initial run with source off
121:                                    source.off()
122:                                    logData(0, settling_time + off_time, pd, fAPerCount[dict['preampConfigs'].index(R)],d,t0)
123:                                    # Run with source on
124:                                    source.on()
125:                                    logData(1, settling_time + on_time, pd, fAPerCount[dict['preampConfigs'].index(R)],d,t0)
126:                                    # Final run with source off
127:                                    source.off()
```

**GenerateData.py**
```
128:                                        logData(0, settling_time + off_time, pd, fAPerCount[
dict['preampConfigs'].index(R)],d,t0)
129:                                print "stop"
130:
131:        sys.stdout=oldstdout
132:        illum.close()
133:        config=open(os.path.join("Runs",RunName,'config.log'),'w')
134:        endTemp=d.getPreampTemp(0)
135:        config.write(WriteConfigFromDict(pd,date,startTemp,endTemp,**dict))
136:        for step,R in enumerate(dict['preampConfigs']):
137:                for x in dict['adcGains']:
138:                        fAPerCount[step]= setupConfigs(preampConfigs[step],pd,x,d)
139:                        config.write("config\trG=%s\t: counts for adcGain %i ch0(1x) and adcGain %i ch1(x101)\n"%(R,x,x))
140:                        config.write("config\tfAPerCount\t%.8g\t%.8g\n"%(fAPerCount[step][pd,0],fAPerCount[step][pd,1]))
141:
142:        config.close()
143:        print "Data successfully taken. Starting to process..."
144:        summarizePD.summarizepd(dirname=os.path.join('Runs', dict["RunName"]))
145:        print "The configfile was successful processed!"
146:
```

**CalculateRatios.py**

```python
1: #!/usr/bin/env python
2:
3: import os
4: from itertools import combinations
5: from numpy import sqrt
6: import matplotlib.pyplot as plt
7: from matplotlib.backends.backend_pdf import PdfPages
8: from copy import deepcopy
9: from Modules import PlotResults
10:
11: ###########Definitions#####################################
12:
13: def get_immediate_subdirectories(a_dir):
14:     return [name for name in os.listdir(a_dir)
15:             if os.path.isdir(os.path.join(a_dir, name))]
16:
17: def gain(key):
18:     val=key[0]*key[1]
19:     return val if key[2]==0 else 101*val
20:
21: def convertToRatio(tocompare):
22:     ratiosToCompare={}
23:     for key1, key2 in combinations(tocompare.keys(), r = 2):
24:         pdcurr=(tocompare[key1][0]/tocompare[key1][1]**2+tocompare[key2][0]/tocompare[key2][1]**2)/(1/tocompare[key1][1]**2+1/tocompare[key2][1]**2)
25:         ratio=tocompare[key1][0]/tocompare[key2][0]

26:         error= sqrt(tocompare[key1][1]**2+tocompare[key2][1]**2) #relative errors
27:         if gain(key1)>gain(key2):
28:             ratiosToCompare[(key1,key2)]=[ratio,ratio*error, pdcurr] #sigma error
29:         else:
30:             ratiosToCompare[(key2,key1)]=[1./ratio,error/ratio, pdcurr] #sigma error
31:     return ratiosToCompare
32:
33: def convertFileToRatio(subdirectory,currentname):
34:     with open(os.path.join("Processed",subdirectory,currentname)) as current:
35:         lines = current.readlines()
36:     lines.pop(0)
37:     tocompare={}
38:
39:     for line in lines:
40:         lineparts=line.split('\t')
41:         if lineparts[4]!="nan":
42:             val0=float(lineparts[4])
43:             relerror0=float(lineparts[5])/val0    #relerror
44: #            if val0>0 and relerror0 < 0.5:
45:             tocompare[int(float(lineparts[2])),int(lineparts[3][4]),0]=[val0,relerror0]
46:         if lineparts[6]!="nan":
47:             val1=float(lineparts[6])
48:             relerror1=float(lineparts[7])/val1        #relerror
49: #            if val1>0 and relerror1 < 0.05:
50:             tocompare[int(float(lineparts[2])),int(lineparts[3][-2]),1]=[val1,relerror1]
51:     return convertToRatio(tocompare)  #returned dict contains sigma errors
52:
53: def mergeRatios(total,current):
54:     for key in current:
55:         valmean=current[key][0]/(current[key][1]**2)
56:         valsigma=1./(current[key][1]**2)
57:         if key in total:
58:             total[key][0]+=valmean
59:             total[key][1]+=valsigma
60:         else:
```

**CalculateRatios.py**
```
 61:                         total[key]=[valmean,valsigma]
 62:         return total
 63:
 64: def updateRatios(ConfigRatios):
 65:         totalRatios={}
 66:         for k in ConfigRatios:
 67:                 totalRatios=mergeRatios(totalRatios,ConfigRatios[k])
 68:         for key in totalRatios:
 69:                 error=totalRatios[key][1]
 70:                 totalRatios[key][0]=(totalRatios[key][0])/error
 71:                 totalRatios[key][1]=sqrt(1./error)
 72:         return totalRatios
 73:
 74: def dictOfDicts(subdirectory,ConfigRatios):
 75:         filelist=os.listdir(os.path.join("Processed",subdirectory))
 76:         for file in filelist:
 77:                 ConfigRatios[file]=convertFileToRatio(subdirectory,file)
 78:         return ConfigRatios
 79:
 80: def consistencyCheck(totalratio,totaldict):
 81:         i=0
 82:         length=0
 83:         badpoints={} # bad points for each file
 84:         badratios={} # bad points for each ratio
 85:         for key in totaldict:
 86:                 errorlist=[]
 87:                 length+=len(totaldict[key])
 88:                 for subkey in totaldict[key]:
 89:                         if subkey in badratios: badratios[subkey][0]+=1
 90:                         else: badratios[subkey]=[1,0,[]]
 91:                         abserror=sigmasingleratio*sqrt(totalratio[subkey][1]**2+totaldict[key][subkey][1]**2)
 92:                         dif=abs(totalratio[subkey][0]-totaldict[key][subkey][0])
 93:                         if dif>abserror:
 94:                                 i+=1
 95:                                 errorlist.append(subkey)
 96:                                 badratios[subkey][1]+=1
 97:                                 badratios[subkey][2].append(key)
 98:                 #for errors in errorlist:
 99:                 #        totaldict[key].pop(errors)
100:                 badpoints[key]=errorlist
101:         print str(i)+" of the " + str(length) + " ratios were not compatible with the averaged mean."
102:         return badpoints, badratios
103:
104:
105: def upgradeRelativeRatio(exact,relative,total):
106:         l=0
107:         combined={}
108:         ConsCheck={}
109:         for key in relative:
110: #                print "###" + str(key)
111:                 if gain(key)>gain(exact):   #upward
112:                         for key2 in total:
113:                                 if key2[1]==key:
114:                                         comratio=relative[key][0]*total[key2][0]
115:                                         comsigma=comratio*sqrt( (relative[key][1]/relative[key][0])**2  + (total[key2][1]/total[key2][0])**2   )
116:                                         tempval=comratio/comsigma**2
117:                                         temperror=1./comsigma**2
118:                                         l+=1
119:
120:
121:                                         if key2[0] in combined:
122: #                                                print "upward add"+ str(key2[0])
123:                                                 combined[key2[0]][0]+=tempval
124:                                                 combined[key2[0]][1]+=temperror
125:                                                 ConsCheck[key2[0]].append((key, comr
```

**CalculateRatios.py**

```
atio, comsigma))
  126:                                   else:
  127: #                                       print "upward new"+ str(key2[0])
  128:                                       combined[key2[0]]=[tempval,temperror]
  129:                                       ConsCheck[key2[0]]=[(key, comratio, comsigma)]
  130:                           else:                      #downward
  131:                               for key2 in total:
  132:                                   if key2[0]==key:
  133:                                       comratio=relative[key][0]/total[key2][0]
  134:                                       comsigma=comratio*sqrt( (relative[key][1]/relative[key][0])**2 +  (total[key2][1]/total[key2][0])**2   )
  135:                                       tempval=comratio/comsigma**2
  136:                                       temperror=1./comsigma**2
  137:                                       l+=1
  138:
  139:                                   if key2[1] in combined:
  140: #                                       print "downward add "+ str(key2[1])
  141:                                       combined[key2[1]][0]+=tempval
  142:                                       combined[key2[1]][1]+=temperror
  143:                                       ConsCheck[key2[1]].append((key, comratio, comsigma))
  144:                                   else:
  145: #                                       print "downward new "+ str(key2[1])
  146:                                       combined[key2[1]]=[tempval,temperror]
  147:                                       ConsCheck[key2[1]]=[(key, comratio, comsigma)]
  148:     print "There are " + str(l) + " ratios combined with one intermediate step."
  149:
  150:     supercombined=deepcopy(relative)
  151:     for k in combined:
  152:         #make the errors in combined correct
  153:         error=combined[k][1]
  154:         combined[k][0]=combined[k][0]/error
  155:         combined[k][1]=sqrt(1./error)
  156:         if k in relative:
  157:             abserror=sqrt(combined[k][1]**2+relative[k][1]**2)
  158:             dif=abs(combined[k][0]-relative[k][0])
  159:             if dif>sigmacombinedratios*abserror: #the direct measurement is kept!
  160:                 print "there was a problem with the ratio between " + str(exact) +" and " + str(k)
  161:             else:
  162:                 mean=combined[k][0]/(combined[k][1]**2)+relative[k][0]/(relative[k][1]**2)
  163:                 err=1./(relative[k][1]*relative[k][1]) + 1./(combined[k][1]*combined[k][1])
  164:                 supercombined[k]=[mean/err,sqrt(1./err)]
  165:         else:
  166:             supercombined[k]=combined[k]
  167:
  168:     return supercombined,combined, ConsCheck
  169:
  170: ################Body################################
  171:
  172:
  173: ###Input###
  174: ratio =tuple(int(x.strip()) for x in raw_input("The exact gain combination is: ").split(','))
  175: check_brightness=raw_input("Make brightness plots? (y/n): ")
  176: check_indirect=raw_input("Make indirect measurement check? (y/n): ")
  177: check_badrawdata=raw_input("Make raw data plots of runs with bad points? (y/n): ")
  178:
  179:
  180: ###Finetune Parameters###
  181: global sigmasingleratio
```

**CalculateRatios.py**

```
182: sigmasingleratio=3
183: global sigmacombinedratios
184: sigmacombinedratios=1.5
185: global ratiobadpoints
186: ratiobadpoints=0.2
187: global numberofbadpointsinfile
188: numberofbadpointsinfile=8
189:
190:
191:
192: ###Printing files and terminal output###
193: print "---------------------------------"
194: dirnames=get_immediate_subdirectories("Processed")
195: ConfigRatios={}
196: for subdirectory in dirnames:
197:         totaldict=dictOfDicts(subdirectory,ConfigRatios)
198:
199: totalRatios=updateRatios(totaldict)
200:
201:
202: # In case deleting bad points from totaldict, make sure you call 'PlotRatios' at this place!
203:
204: badpoints, badratios=consistencyCheck(totalRatios,totaldict)
205: totalRatios=updateRatios(totaldict)
206: #badpoints=consistencyCheck(totalRatios,totaldict)
207:
208: # Before deleting the files with more than 'ratiobadpoints' % of the total number of data being bad, we will make a deep copy of all data, used for plotting at the end.
209: totRatCopy=deepcopy(totalRatios)
210:
211:
212: if not os.path.exists('Results/Bad'): os.makedirs('Results/Bad')
213:
214: badfiles=[]
215: with open(os.path.join("Results","Bad","BadPoints.txt"), "w") as f:
216:         total,l,m=0,0,0
217:         f.write('Relative number of bad points for each ratio:\n')
218:         for key in badratios:
219:                 if float(badratios[key][1])/badratios[key][0]>ratiobadpoints or badratios[key][0]<2:
220:                         f.write("(%s,%s,%s), (%s,%s,%s):%s/%s\n"%("{:.0e}".format(key[0][0]),key[0][1], key[0][2], "{:.0e}".format(key[1][0]),key[1][1], key[1][2],badratios[key][1],badratios[key][0]))
221:                         for i in range(len(badratios[key][2])):
222:                                 f.write("\t%s\n"%(sorted(badratios[key][2], key=lambda x: totaldict[x][key][2] )[i]))
223:                         total+=badratios[key][1]
224:                         if badratios[key][0]<2: m+=1
225:                         else: l+=1
226:                         #if badratios[key][1]!=badratios[key][0]:
227:                         totalRatios.pop(key)
228:
229:         print str(total-m)+" bad points were associated with the "+ str(l) + " ratios, which are excluded because more than " +str(ratiobadpoints*100)+"% of total number of points were bad/red."
230:         print str(m) + " ratios, which each contained 1 point were excluded."
231:         print "---------------------------------"
232:         f.write("\n\n\n\n")
233:
234:         for (key, value) in badpoints.iteritems():
235:                 f.write("%s\n"%(key))
236:                 count=0
237:                 for i in range(len(value)):
238:                         count+=1
239:                         f.write("(%s,%s,%s), (%s,%s,%s)\n"%("{:.0e}".format(value[i][0][0]), value[i][0][1], value[i][0][2], "{:.0e}".format(value[i][1][0]), value[i][1][1], value[i][1][2]))
```

**CalculateRatios.py**
```
240:                         f.write("\n")
241:                 if count>numberofbadpointsinfile: badfiles.append(key)
242:
243: with open(os.path.join("Results","Ratios.txt"), "w") as f:
244:         f.write("config1 \t\t\t\t config2 \t\t\t\t\t\t ratio \t\t sigma\n")
245:         for (key, val) in totalRatios.iteritems():
246:                 f.write("%-*s%-*s"%(25,key[0],20,key[1]))
247:                 f.write("\t\t\t")
248:                 f.write("%-*.4f\t%-*.4f"%(10,val[0],10,val[1]))
249:                 f.write("\n")
250:
251: print "There are " + str(len(totalRatios)) + " calculated ratios."
252:
253: with open(os.path.join("Results", "Bad", "BadToDo.mcr"), "w") as todo:
254:         for i in range(len(badfiles)):
255:                 todo.write("%s\n"%('_'.join(badfiles[i].split('_')[0:4])+'.config'))
256:
257: j=0
258: # Return number of ratios which are precise with 0.1%
259: for (key, val) in totalRatios.iteritems():
260:         if val[1]<0.0011: j+=1
261: print "There are " + str(j) + " calculated ratios with 0.1% precision."
262:
263: i=0
264: relativeRatio={}
265: for key in totalRatios:
266:         if key[0]==ratio:
267:                 i+=1
268:                 relativeRatio[key[1]]=[1./totalRatios[key][0],totalRatios[key][1]/(totalRatios[key][0]**2)]
269:         elif key[1]==ratio:
270:                 i+=1
271:                 relativeRatio[key[0]]=totalRatios[key]
272:
273: print "There are " +str(i)+ " direct ratios found that compare with " + str(ratio) + "."
274: superCombinedRatio,combinedRatio, consCheck=upgradeRelativeRatio(ratio,relativeRatio,totalRatios)
275:
276: k=0
277: with open(os.path.join("Results","Comparedto%s.txt"%(str(ratio))), "w") as f:
278:         f.write("### SuperCombined ###")
279:         f.write("\n")
280:         f.write("configuration \t\t ratio \t sigma\n\n")
281:         for (key, val) in superCombinedRatio.iteritems():
282:                 f.write("%-*s"%(25,key))
283:                 f.write("%-*.5f\t%-*.5f"%(10,val[0],10,val[1]))
284:                 f.write("\n")
285:                 k+=1
286:         f.write("\n")
287:         f.write("\n")
288:
289:         f.write("### Combined ###")
290:         f.write("\n")
291:         f.write("configuration \t\t ratio \t sigma\n\n")
292:         for (key, val) in combinedRatio.iteritems():
293:                 f.write("%-*s"%(25,key))
294:                 f.write("%-*.5f\t%-*.5f"%(10,val[0],10,val[1]))
295:                 f.write("\n")
296:
297:         f.write("\n")
298:         f.write("\n")
299:         f.write("### Direct ###")
300:         f.write("\n")
301:         f.write("configuration \t\t ratio \t sigma\n\n")
302:         for (key, val) in relativeRatio.iteritems():
303:                 f.write("%-*s"%(25,key))
304:                 f.write("%-*.5f\t%-*.5f"%(10,val[0],10,val[1]))
```

**CalculateRatios.py**
```
305:                    f.write("\n")
306:
307: print "There are " +str(k)+ " ratios found that compare with " + str(ratio) +"."
308: print "-------------------------------"
309:
310:
311:
312: ###Plotting###
313:
314: if check_brightness=='y': PlotResults.PlotRatios(totRatCopy, totaldict, sigmasingleratio)
315:
316: if check_indirect=='y':
317:         print "Indirect Measurement plotting:"
318:         PlotResults.PlotConsist(relativeRatio, superCombinedRatio, combinedRatio, consCheck, 1.5)
319: m=1
320: if check_badrawdata=='y':
321:         print "Bad raw data plotting:"
322:         for i in range(len(badfiles)):
323:                 print "Plot: "+str(m)+"/"+str(len(badfiles))
324:                 PlotResults.PlotRawData('.'.join(badfiles[i].split('.')[0:len(badfiles[i].split('.'))-1]))
325:                 m+=1
326:
327:
328:
329:
```



\section{Acknowledgement}
I would like to thank my colleague Jeroen Van Houtte for developing most of the graphical output of the program and for his valuable contributions to all parts of the end result. Further I thank my supervisor Chris Wendt who came up with illuminating ideas when they were needed the most. Finally, this \textit{Honours Award in Sciences} would never looked the same without  professor Dirk Ryckbosch. He gave me the opportunities to improve my skills while collaborating in research of the highest level and was there with advice whenever obstacles appeared. 
\nocite{*}
\bibliography{IceCube}
\bibliographystyle{unsrt}

\end{document}